\documentclass[twocolumn,pre,showpacs,preprintnumbers,amsmath,amssymb,floatfix,superscriptaddress]{revtex4-1}
\usepackage{graphicx}
\usepackage{amsmath}
\usepackage{amssymb}
\usepackage{color}

\begin{document}

\title{Impact of tunneling anisotropy on the conductivity of nanorod dispersions}

\author{Biagio Nigro}\affiliation{LPM, Ecole Polytechnique F\'ed\'erale de
Lausanne, Station 17, CP-1015 Lausanne, Switzerland}
\author{Claudio Grimaldi}\email{claudio.grimaldi@epfl.ch}\affiliation{Laboratory of Physics of Complex Matter, 
Ecole Polytechnique F\'ed\'erale de Lausanne, Station 3, CP-1015 Lausanne, Switzerland}

\begin{abstract}
While the tunneling conductance between two spherical-like conducting particles depends on the relative inter-particle 
distance, the wave function overlap between states of two rod-like particles, and so the tunneling conductance, depends also on 
the relative orientation of the rod axes. Modeling slender rod-like particles as cylindrical quantum wells of diameter
$D$ and length $L\gg D$, we calculate the matrix element of the tunneling between two rods for arbitrary relative
orientations of the rod axes. We show that tunneling between two parallel rods is about $L/\sqrt{D\xi}$ times larger 
than the tunneling matrix element for perpendicular rods, where $\xi$ is the tunneling decay length.
By considering the full dependence of the tunneling conductance on the angle between rod axes,
we calculate within an effective medium theory the conductivity of dispersions of rods with different degrees of alignment. 
We find that for isotropically oriented rods, the effect of orientation in the tunneling processes is marginal for all rod concentrations. 
On the contrary, for systems of strongly aligned rods, the enhanced tunneling between nearly parallel rods increases significantly
the system conductivity in a relatively large concentration range. Next,
we consider systems in which short-range attraction between rods is added, as in dispersions of rods with depletion interaction. 
We find that the strongly anisotropic attraction promotes enhanced tunneling between neighboring parallel rods,
increasing the effective medium conductivity by several orders of magnitude compared to the case in which the angular
dependence of tunneling is ignored, even for relatively weak attractions.

\end{abstract}


\maketitle

\section{Introduction}
\label{intro}

The electrical transport properties of nanocomposites are potentially enhanced by conducting
fillers with high aspect-ratios, as for particles having rod-like or disk-like geometries. 
For examples, polymeric composites containing carbon nanotubes
or conducting nanofibers display relatively large conductivities even at particle loadings below 
one percent in volume fraction \cite{Gelves2006,Bauhofer2009,Al-Saleh2009}. At such small loadings the physical properties of the insulating
medium are preserved, leading for example to electrically conducting composites which are also lightweight 
and mechanically flexible. 

In dispersions of conducting nanofillers, the electrical connectivity is established by
quantum tunneling or hopping of electrons through the insulating regions separating the fillers.
The enhanced conductivity at small loading of high-aspect-ratio fillers is basically
explained by the large excluded volume associated to the fillers, which has the net effect
of reducing the filler separation needed to establish tunneling connectivity \cite{Balberg1984,Garboczi1995,Celzard1996}.
Excluded volume arguments predicts also that anisotropy in the orientation of the fillers
decreases the system conductivity \cite{Winey2009}, as mutual alignment of rod-like fillers
reduces the excluded volume at a given loading.

Theories and computer simulations, either using the percolation approach to describe the
electrical connectedness \cite{Balberg1984,Bug1985,Neda1999,Foygel2005,Schilling2007,Berhan2007,Kyrylyuk2008,Chatterjee2008,Otten2012} 
or considering explicitly its tunneling/hopping nature \cite{Ambrosetti2010a,Nigro2013c}, confirm
the general trend predicted by the excluded volume argument. Further theoretical studies have
investigated the role of filler waviness \cite{Chatterjee2002,Dalmas2006,Berhan2007b,Li2008}, 
attractive forces between the fillers \cite{Schilling2007,Kyrylyuk2008,Nigro2013c}, and
filler size polydispersity \cite{Otten2009,Chatterjee2010,Mutiso2012,Nigro2013a} on the electrical connectedness 
in systems of high-aspect ratio particles, extending thus our understanding of this class of 
nanocomposizes \cite{Schilling2010,Mutiso2014}.

However, the current state of research
is based on the assumption that the electrical connectivity between any two high-aspect-ratio fillers depends
only on their relative distance, and ignores possible contributions arising from the relative orientation of
the fillers. One expects instead that the probability of electron tunneling between two perpendicular rods
is smaller than that occurring between two perfectly aligned rods at the same distance. 
This is so because the overlap between the wave functions centered on two parallel rods extends over the whole length 
of the rods, while for perpendicular rods the overlap is limited to the region of closest approach.

An interesting question is thus whether the orientation dependence of tunneling has any relevant effect on the 
conductivity of composites with rod-like fillers. In particular, clarifying how enhanced tunneling 
between aligned rods competes with the reduction of connectivity of anisotropic rod orientations may be of 
special relevance for those nanotube and nanofiber composites with high degrees of rod alignment \cite{Du2005,Wang2008}.
Even more compelling is the issue concerning the role of tunneling anisotropy on the conductivity
of rods experiencing van der Waals forces or depletion interactions \cite{Vigolo2005,Maillaud2013}, as these induce 
a strongly anisotropic attraction that favors alignment of neighboring rods \cite{Kyrylyuk2008,Schoot1992,Odijk1994,Lekkerkerker1997,Schiller2010}.

In this paper, we derive the dependence of the tunneling matrix element on the 
relative orientation of two slender conducting rods. We show that the resulting tunneling conductance strongly
increases as the angle between the two rod axes decreases, and that it decays 
exponentially with the shortest distance between the two rod cores. By using the full functional 
dependence of tunneling on the spatial configuration of the rods, we calculate the composite conductivity within an effective medium approximation
for dispersions of rods with high aspect-ratio. We show that for isotropically distributed rod orientations, the angular dependence
of tunneling has only a marginal effect for all volume fractions, confirming that tunneling transport in this case is dominated
solely by the relative distances between the rods. On the contrary, for dispersions with high-degrees of rod alignments, we find that anisotropic 
tunneling may enhance significantly the effective medium conductivity with respect to the case in which the angular
dependence of tunneling is ignored. We also consider suspensions of rods with short-range attractive interactions, as to
simulate the effective attraction that rods experience when small depletant particles are added to the system. 
We show that the tunneling matrix element and the attraction potential between the rods have a similar 
angular dependence and that they combine together to increase the contribution of aligned rods to the composite 
conductivity. We find that in the presence of anisotropic tunneling even moderate attractions can enhance the effective 
conductivity by several orders of magnitude.

\section{tunneling matrix element}
\label{sec:matrix}

To describe tunneling of electrons between rod-like nanoparticles, we model the particle geometry by a cylinder
of length $L$ and diameter $D$, and consider each cylinder as a quantum well in which electrons are confined by 
a square-well potential $U(\mathbf{r})$. In the following, we shall restrict our analysis to the
tunneling conductance between the lateral surfaces of two cylinders, as the shortest interparticle distance
in dispersions of cylinders with large $L/D$ is predominantly between their respective axes.
In this limit, we neglect details of the confining potential at the cylinder ends, and consider it as being
given by infinite hard walls.
For an isolated cylinder centered at the origin and with its main axis directed along $z$ we thus take 
$U(\mathbf{r})=U_\rho(\rho)+U_z(z)$, with
\begin{equation}
\label{sw}
U_\rho(\rho)=\left\{
\begin{array}{ll}
0, & \rho\leq R,\\
U_0>0, & \rho > R,
\end{array}\right.
\end{equation}
where $R=D/2$ and $\rho$ is the radial distance from the cylinder axis, and
\begin{equation}
\label{Uz}
U_z(z)=\left\{
\begin{array}{ll}
0, & \vert z\vert\leq L/2,\\
\infty, & \vert z\vert > L/2.
\end{array}\right.
\end{equation}
With this form of the confining potential, the stationary Schr\"odinger equation
\begin{equation}
\label{sch1}
-\frac{\hbar^2}{2m_e}\nabla^2\psi(\mathbf{r})+U(\mathbf{r})\psi(\mathbf{r})=E\psi(\mathbf{r}),
\end{equation}
is separable in cylindrical coordinates. 
The bound states solution of Eq.~\eqref{sch1} have the form \cite{LeGoff1993,Li2001,Liu2008}:
\begin{equation}
\label{psi1}
\psi(\mathbf{r})=\left\{
\begin{array}{lll}
a J_m\!\left(\rho\frac{\sqrt{2m_eE^r_m}}{\hbar}\right)f_k(z)e^{im\phi} & \textrm{for} & \rho\leq R,\\
b K_m\!\left(\rho\frac{\sqrt{2m_e(U_0-E^r_m)}}{\hbar}\right)f_k(z)e^{im\phi} & \textrm{for} & \rho > R,
\end{array}\right.
\end{equation}
with eigenvalue $E=E^r_m+E^z_k$, where $E^r_m$ are the energy levels for the radial motion (with $m=0,\, \pm 1,\,\pm 2,\ldots$), 
and $E^z_k=\hbar^2 k^2/2m_e$ are the energy levels for the motion along $z$ and $k=\pi n/L$ (with $n=1,2,3,\ldots$).
In Eq.~\eqref{psi1}, $a$ and $b$ are constants, $J_m$ and $K_m$ are respectively Bessel functions and
modified Bessel functions, $\phi$ is the azimuthal angle on the $xy$-plane, and
\begin{equation}
\label{f}
f_k(z)=\sqrt{\frac{2}{L}}\sin\left[k(z+L/2)\right]\theta(L/2-\vert z\vert),
\end{equation}
is the one dimensional wave function for one electron in the segment $-L/2\leq z\leq L/2$,
and $\theta(x)=1$ for $x\geq 0$ and $\theta(x)=0$ for $x<0$. 
The energy levels $E^r_m$ can be found by imposing boundary conditions on the cylinder surface. 
We note however that for $L/D\gg 1$ the much stronger confinement in the radial direction as
compared to the $z$ direction allows us to assume that only the lowest electronic sub-band in 
the radial direction is occupied. We thus take $m=0$ in Eq.~\eqref{psi1} and define the wave
function outside the cylindrical well as:
\begin{equation}
\label{psi2}
\psi_k(\mathbf{r})=\frac{\varphi(R)}{K_0(R/\xi)}K_0(\rho/\xi)f_k(z) \,\,\,\textrm{for}\,\,\,\rho>R,
\end{equation}
where
\begin{equation}
\label{kappa}
\xi=\hbar/\sqrt{2m_e(U_0-E^r_0)}
\end{equation}
is the tunneling decay length and $\varphi(R)=a J_0\!\left(R\sqrt{2m_eE^r_0}/\hbar\right)$ is the radial wave function on the 
cylindrical surface. Noting that $K_0(x)\simeq \sqrt{\pi/2x}\exp(-x)$ for $x\gg 1$, we see that in the strong 
localization regime  $\xi/R \ll 1$ the wave function \eqref{psi2} falls off exponentially with the distance $\rho-R$
from the cylinder surface, that is: $\psi_k(\mathbf{r})\propto\exp[-(\rho-R)/\xi]$ for $\rho-R\geq 0$.

We proceed by considering two identical cylinders, denoted hereafter as $i$ and $j$, centered at
$\mathbf{r}_i$ and $\mathbf{r}_j$ and having their axes oriented arbitrarily.
We assume also that the two cylinders do not overlap each other and that a weak potential drop
$V$ is applied between them. At low temperatures,
the tunneling current flowing across the region separating $i$ and $j$ is \cite{Tersoff1985,Bardeen1961,Gottlieb2006}:
\begin{equation}
\label{curr}
I\simeq \frac{2\pi e^2 V}{\hbar}\sum_{k,k'}
\vert M_{i k,j k'}\vert^2\delta(E_F-E^z_k)\delta(E_F-E^z_{k'}),
\end{equation}
where $e$ is the electron charge, $M_{ik,jk'}$ is the tunneling matrix element 
between states $\psi_{ik}$ and $\psi_{jk'}$ localized on $i$ and $j$,
and $E_F$ is the Fermi energy measured with respect to $E^r_0$.
The two $\delta$-functions in Eq.~\eqref{curr} restrict the wave numbers $k$ and $k'$
to $k_F=\sqrt{2m_e E_F}/\hbar$, and the tunneling current reduces thus to:
\begin{equation} 
\label{curr2}
I\simeq \frac{2\pi e^2 V}{\hbar}(LN_F)^2\vert M_{i,j}\vert^2,
\end{equation}
where $M_{i,j}=M_{i k_F,j k_F}$ and $N_F=(1/L)\sum_k\delta(E_F-E^z_k)\simeq m_e/(\pi\hbar^2 k_F)$ 
is the density of states at the Fermi level for orbits along the cylinder axis. We note that since the
energy levels for the radial motion scale approximately as $\hbar^2/(2m_e R^2)$,
the condition that only the $m=0$ state is occupied is satisfied by assuming $k_F R<1$. 

Following Bardeen's formalism \cite{Bardeen1961,Gottlieb2006}, we express $M_{i,j}$
as an integral over the surface $\Sigma$ lying entirely within the region separating
the two cylinders:
\begin{equation}
\label{tun1}
M_{i,j}=-\frac{\hbar^2}{2m_e}\int_\Sigma d\boldsymbol{s}\cdot
\left[\psi_{j}(\mathbf{r})^*\boldsymbol{\nabla} 
\psi_{i}(\mathbf{r})-\psi_{i}(\mathbf{r})\boldsymbol{\nabla}\psi_{j}(\mathbf{r})^*\right],
\end{equation}
where $d\boldsymbol{s}$ is the differential vector normal to $\Sigma$, $\psi_{i}=\psi_{i k_F}$,
and $\psi_{j}=\psi_{j k_F}$.

\begin{figure}[t]
\begin{center}
\includegraphics[scale=0.38,clip=true]{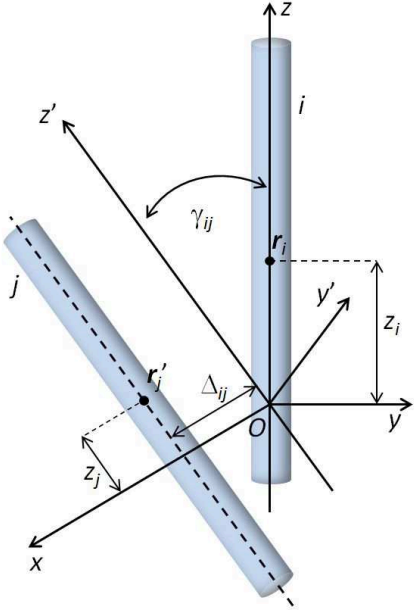}
\caption{(Color online) Schematic representation of two cylinders, $i$ and $j$, whose
axes are tilted by an angle $\gamma_{ij}$. The axis of cylinder $i$ is directed along $z$,
while that of $j$ is directed along $z'$. The axis $x$ is directed along the shortest distance,
of length $\Delta_{ij}$, between $i$ and $j$. $z_i$ and $z_j$ are the positions of the
centers of mass of $i$ and $j$ along their respective axes.}
\label{fig1}
\end{center}
\end{figure}

Since the separation surface is outside the cores of the two cylindrical wells, we express 
$\psi_{i}$ and $\psi_{j}$ in terms of the wave function \eqref{psi2} in the barrier region. 
To this end, we consider the configuration depicted in Fig.~\ref{fig1}, in which the axes 
of $i$ and $j$ are tilted by an angle $\gamma_{ij}$ and the shortest line between $i$ and $j$, 
of length $\Delta_{ij}$, connects their respective axes. 
We introduce an orthogonal coordinate system $Oxyz$ having
$z$-axis along the axis of cylinder $i$ and $x$-axis along the shortest line connecting $i$ and $j$.
A second coordinate system, $Oxy'z'$, has origin in $O$ and is rotated with respect to $Oxyz$ 
by an angle $\gamma_{ij}$ about the $x$-axis, as shown in Fig.~\ref{fig1}.
The position vectors $\mathbf{r}$ in $Oxyz$ and $\mathbf {r}'$ in the rotated frame $Oxy'z'$
are thus related to each other by $\mathbf{r}'=\hat{\mathbf{R}}_{ij}\mathbf{r}$, where 
\begin{equation}
\label{Rot}
\hat{\mathbf{R}}_{ij}=\left(
\begin{array}{ccc}
1 & 0 & 0  \\
0 & \cos\gamma_{ij} & \sin\gamma_{ij}  \\
0 & -\sin\gamma_{ij} & \cos\gamma_{ij}
\end{array}\right),
\end{equation} 
is the matrix for the rotation of an angle $\gamma_{ij}$ about the $x$-axis.
Denoting the position of the centers of mass of $i$ and $j$ along their
respective axes by $z_i$ and $z_j$, where $-L/2\leq z_i,z_j\leq L/2$ (see Fig.~\ref{fig1}), 
we write $\psi_i$ and $\psi_j$ as:
\begin{align}
\label{psi3}
\psi_{i}&=\psi_{k_F}(\mathbf{r}-\mathbf{r}_i), \\
\label{psi4}
\psi_{j}&=\psi_{k_F}(\mathbf{r}'-\mathbf{r}'_j)=\psi_{k_F}(\hat{\mathbf{R}}_{ij}\mathbf{r}-\mathbf{r}'_j),
\end{align}
where 
\begin{equation}
\label{r}
\mathbf{r}_i=\left(
\begin{array}{c}
0 \\
0 \\
z_i
\end{array}\right),\,\,\,\,
\mathbf{r}'_j=\left(
\begin{array}{c}
\Delta_{ij} \\
0 \\
z_j
\end{array}\right).
\end{equation}

To solve the surface integral in Eq.~\eqref{tun1}, we generalize the Green's function method of 
Ref.~\cite{Chen1990}. To this end, we consider the equation for the Green's function
associated to a line source on the $z$-axis \cite{ArfkenBook}: 
\begin{equation}
\label{green1}
(\nabla^2 -\xi^{-2})G(\boldsymbol{\rho})=-\delta(\boldsymbol{\rho}),
\end{equation}
where $\boldsymbol{\rho}$ is the radial vector on the $xy$-plane. Noting that the solution of Eq.~\eqref{green1} which
is regular at $\rho\rightarrow\infty$ is $G(\boldsymbol{\rho})=K_0(\rho/\xi)/2\pi$, and using Eq.~\eqref{psi3},
we can express the wave function of $i$ outside the core as:
\begin{equation}
\label{psi5}
\psi_i=\frac{2\pi\varphi(R)}{K_0(R/\xi)}G(\boldsymbol{\rho})f_{k_F}(z-z_i).
\end{equation}
We thus rewrite Eq.~\eqref{tun1} as:
\begin{align}
\label{tun2}
M_{ij}=-C\int_\Sigma d\boldsymbol{s} \cdot &
\left\{\psi_{k_F}(\mathbf{r}'-\mathbf{r}'_j)\boldsymbol{\nabla}\left[G(\boldsymbol{\rho})f_{k_F}(z-z_i)\right]\right. \nonumber \\
&-\left.G(\boldsymbol{\rho})f_{k_F}(z-z_i)\boldsymbol{\nabla}\psi_{k_F}(\mathbf{r}'-\mathbf{r}'_j)\right\},
\end{align}
and use the divergence theorem to convert the surface integral into an integral over the volume $\Omega_i$ 
which contains the cylinder $i$:
\begin{align}
\label{tun3}
M_{ij}=C\int_{\Omega_i} \!d\mathbf{r} & 
\left\{G(\boldsymbol{\rho})f_{k_F}(z-z_i)\nabla^2\psi_{k_F}(\mathbf{r}'-\mathbf{r}'_j)\right. \nonumber \\
&-\left.\psi_{k_F}(\mathbf{r}'-\mathbf{r}'_j)\nabla^2\left[G(\boldsymbol{\rho})f_{k_F}(z-z_i)\right]\right\},
\end{align}
where
\begin{equation}
\label{c}
C=\frac{\pi\hbar^2}{m_e}\frac{\varphi(R)}{K_0(R/\xi)}.
\end{equation}
Using $\nabla^2\psi_{k_F}(\mathbf{r}'-\mathbf{r}'_j)=\nabla'^2\psi_{k_F}(\mathbf{r}'-\mathbf{r}'_j)
=(\xi^{-2}-k_F^2)\psi_{k_F}(\mathbf{r}'-\mathbf{r}'_j)$, which results from the rotational invariance of $\nabla^2$, and
\begin{align}
\nabla^2\left[G(\boldsymbol{\rho})f_{k_F}(z-z_i)\right]=&(\xi^{-2}-k_F^2)G(\boldsymbol{\rho})f_{k_F}(z-z_i)\nonumber \\
&-\delta(\boldsymbol{\rho})f_{k_F}(z-z_i).
\end{align}
which comes from Eq.\eqref{green1}, we find that $M_{ij}$ reduces to:
\begin{align}
\label{tun4}
M_{ij}=C\int_{\Omega_i} &\!d\mathbf{r}\,\psi_{k_F}(\mathbf{r}'-\mathbf{r}'_j)f_{k_F}(z-z_i)\delta(\boldsymbol{\rho}) \nonumber \\
=\frac{C\varphi(R)}{K_0(R/\xi)}\int_{-\infty}^{+\infty}\! &dz\, f_{k_F}(z-z_i)f_{k_F}(z\cos\gamma_{ij}-z_j) \nonumber \\
&\times K_0\!\left(\frac{1}{\xi}\sqrt{\Delta_{ij}^2+z^2\sin^2\gamma_{ij}}\right),
\end{align}
where we have used $\delta(\boldsymbol{\rho})=\delta(x)\delta(y)$. From Eqs.~\eqref{curr2}, \eqref{c}, and 
\eqref{tun4} we thus obtain that the tunneling conductance $g_{ij}=I/V$ is
\begin{equation}
\label{tun6}
g_{ij}=\frac{2\pi e^2}{\hbar}\left[\frac{D\varphi(R)^2}{k_F}\right]^2\vert I_{ij}\vert^2\equiv g_0\vert I_{ij}\vert^2,
\end{equation}
where $g_0$ is a conductance prefactor, $\vert I_{ij}\vert^2=I_{ij}I_{ji}$, and
\begin{align}
\label{i0}
I_{ij}=\frac{L}{K_0(R/\xi)^2D}\int_{-\infty}^{+\infty}\! &dz\, f_{k_F}(z-z_i)f_{k_F}(z\cos\gamma_{ij}-z_j) \nonumber \\
&\times K_0\!\left(\frac{1}{\xi}\sqrt{\Delta_{ij}^2+z^2\sin^2\gamma_{ij}}\right).
\end{align}
To assess the dominant contribution to Eq.~\eqref{i0} of the relative position and orientation of $i$ and $j$,
we ignore the wave modulation along the cylinder axes and replace $f_{k_F}(z)$ with
$\sqrt{1/L}\theta(L/2-\vert z\vert)$, where $\sqrt{1/L}$ is the normalization factor: 
\begin{widetext}
\begin{equation}
\label{i1}
\vert I_{ij}\vert\simeq\frac{\displaystyle\int_{-\infty}^{+\infty}\! dz\, \theta(L/2-\vert z-z_i\vert)
\theta(L/2-\vert z\cos\gamma_{ij}-z_j\vert)
K_0\!\left(\frac{1}{\xi}\sqrt{\Delta_{ij}^2+z^2\sin^2\gamma_{ij}}\right)}
{\displaystyle K_0(R/\xi)^2D}.
\end{equation}
\end{widetext}
We compare the above expression with Eq.~\eqref{i0} in the Appendix, where we show that Eq.~\eqref{i1} 
is approximately equivalent to consider $\vert I_{ij}\vert$ as given by the envelope of 
the wave functions for the motion along $z$. 

Although a general analytical solution of the integral over $z$ is not possible, Eq.~\eqref{i1}
admits simple expressions when the axes are parallel ($\gamma_{ij}=0$) or highly skewed 
($D/L\ll \gamma_{ij}\leq \pi/2$).
For the case $\gamma_{ij}=0$ we find:
\begin{align}
\label{i2}
\vert I_{ij}^\parallel\vert&\simeq\frac{K_0(\Delta_{ij}/\xi)}{DK_0(R/\xi)^2}\int_{-L/2}^{+L/2}\! dz\, 
\theta(L/2-\vert z+z_i-z_j\vert)\nonumber \\
&\simeq (L-\vert z_i-z_j\vert)\frac{e^{-(\vert\Delta_{ij}\vert-D)/\xi}}{\sqrt{2\pi D\xi}},
\end{align}
where we have used the asymptotic limit $K_0(x)\sim \sqrt{\pi/2x}\exp(-x)$. 
Since we are considering $\xi/D\ll 1$, we have also set $\vert\Delta_{ij}\vert\simeq D$
in the pre-exponential factor.

For highly skewed cylinders ($D/L\ll \gamma_{ij}\leq \pi/2$), we note that the exponentially decay of $K_0$
limits the dominant contribution of the $z$ integration to $z\lesssim \Delta_{ij}\ll L$, so that for
$\vert z_i\vert,\vert z_j\vert < L/2$ the two $\theta$-functions in \eqref{i1} are simply unity.
The integration can be performed analytically \cite{GradshteynBook}, leading to:
\begin{align}
\label{i3}
\vert I_{ij}^\perp\vert&\simeq \frac{1}{DK_0(R/\xi)^2}
\int_{-\infty}^{+\infty}\!dz\,
K_0\!\left(\frac{1}{\xi}\sqrt{\Delta_{ij}^2+z^2\sin^2\gamma_{ij}}\right)\nonumber \\
&=\frac{\pi\xi e^{-\vert\Delta_{ij}\vert/\xi}}{\vert \sin\gamma_{ij}\vert DK_0(R/\xi)^2}
\simeq \frac{e^{-(\vert\Delta_{ij}\vert-D)/\xi}}{\vert \sin\gamma_{ij}\vert}.
\end{align}

\begin{figure}[t]
\begin{center}
\includegraphics[scale=0.64,clip=true]{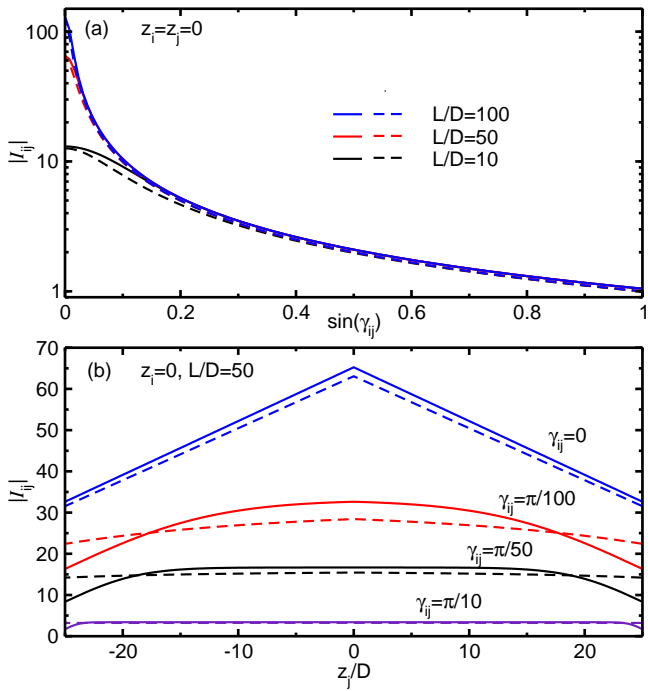}
\caption{(Color online) Comparison between the dimensionless tunneling matrix element $\vert I_{ij}\vert$ of
Eq.~\eqref{i1} (solid lines) and the analytical formula of Eq.~\eqref{iapp} (dashed lines).
The cylinders are at contact ($\Delta_{ij}=D$) and the tunneling decay length is fixed at $\xi=0.1D$.
(a): $\vert I_{ij}\vert$ as a function of $\sin\gamma_{ij}$ for cylinders with aligned centers of mass ($z_i=z_j$)
and different aspect-ratios $L/D$. (b): $\vert I_{ij}\vert$ as a function of $z_j$ for $z_i=0$, $L/D=50$, 
and different angles $\gamma_{ij}$. }\label{fig2}
\end{center}
\end{figure}

From Eqs.~\eqref{i2} and \eqref{i3} we see that the exponential decay of the tunneling matrix element
is unaffected by the relative axis orientation, while for $z_i=z_j$ the pre-exponential factor for parallel 
cylinders is about  $L/\sqrt{D\xi}\gtrsim L/D$ times larger than that for perpendicular cylinders. 
Clearly, the enhanced $\vert I_{ij}\vert$ for small $\gamma_{ij}$ stems from the overlap of the wave 
functions which extends over the whole length of the cylinders when they are parallel, while perpendicular 
or highly skewed cylinders have a much smaller region of overlap. For parallel cylinders with misaligned
centers of mass, the reduction of the wave function overlap for $z_i\neq z_j$ is automatically taken into 
account by the factor $(L-\vert z_i-z_j\vert )$ in Eq.~\eqref{i2}. 

Equations \eqref{i0} and \eqref{i1} are of limited practical use in studying the effect of tunneling anisotropy in
systems of conducting cylinders, as the dependence of $\vert I_{ij}\vert$ on $\Delta_{ij}$, $\gamma_{ij}$,
and on $z_i$ and $z_j$ can be evaluated only after performing a numerical integration over $z$. However,
using the limiting behaviors of Eqs.~\eqref{i2} and \eqref{i3}, we can approximate $\vert I_{ij}\vert$ by 
the following analytical formula:
\begin{equation}
\label{iapp}
\vert I_{ij}\vert\approx\frac{L-\vert z_{ij}\vert}
{\sqrt{2\pi D\xi+(L-\vert z_{ij}\vert)^2\sin^2\!\gamma_{ij}}}e^{-(\vert\Delta_{ij}\vert-D)/\xi},
\end{equation}
where $z_{ij}=z_i-z_j$. Equation \eqref{iapp}
reproduces quite accurately the $\gamma_{ij}$ dependence of Eq.~\eqref{i1}, as shown in Fig.~\ref{fig2}(a)
for the case $z_i=z_j=0$ and $\xi/D=0.1$. The effect of misalignment of the centers of mass is shown 
in Fig.~\ref{fig2}(b), where Eq.~\eqref{i1} (solid lines) and Eq.~\eqref{iapp} (dashed lines) are plotted
by varying $z_j$ from $-L/2$ to $L/2$ and for $z_i=0$ and different $\gamma_{ij}$ values. 
The simple analytical expression in Eq.~\eqref{iapp} captures thus the essential dependence of the
tunneling matrix element on the tunneling variables, and permits a systematic study of the tunneling
anisotropy effects with limited computational effort.

\section{Effective medium approximation for the conductivity}
\label{sec:ema}

To evaluate the effect of tunneling anisotropy on the conductivity of dispersions of 
conducting cylinders we employ the two-site effective medium approximation 
(EMA) \cite{Ambrosetti2010b,Grimaldi2011,Grimaldi2014}. 
For a general system of $N$ conducting particles that are electrically connected through 
tunneling processes, this method amounts to construct a tunneling resistor
network where each node of the network represents a conducting particle and where any pair of nodes,
for example $i$ and $j$, are connected by a tunneling conductance $g_{ij}$. The resulting
network is a complete graph with $N(N-1)/2$ bonds.
EMA consists in finding the effective conductance $\bar{g}$ between any two nodes such that when all
$g_{ij}$'s are replaced by $\bar{g}$, the average resistance of the network remains the same.
It can be shown \cite{Ambrosetti2010b,Grimaldi2011,Grimaldi2014} that the EMA conductance  
$\bar{G}=N\bar{g}/2$ of the effective network within the two-site approximation is the solution of the 
following equation:
\begin{equation}
\label{ema1}
\frac{1}{N}\left\langle\sum_{i,j}{}'\frac{g_{ij}}{g_{ij}+\bar{G}}\right\rangle=2,
\end{equation}
where $\langle\cdots\rangle$ indicates an ensemble average over configurations and
the prime symbols means that $i=j$ is omitted from the summation. $\bar{G}$ is independent of
the system size, and can be considered as a measure of the system conductivity averaged over
all directions.
A short derivation of Eq.~\eqref{ema1} is outlined in Refs.~\cite{Ambrosetti2010b,Grimaldi2014}, while a 
more general description of the method can be found in Ref.~\cite{Grimaldi2011}. 
Relevant features of the two-site tunneling EMA are that it relates explicitly the network conductance with
the structure of the conducting particle dispersions and that it provides very good agreements
with conductivities calculated from simulations of many different systems, as fluids of hard-core \cite{Ambrosetti2010b} or
attractive conducting spheres \cite{Nigro2012a}, segregated distributions of spheres \cite{Nigro2012b}, and fluid mixtures 
of conducting hard-core spherocylinders and insulating spherical depletants \cite{Nigro2013c}.

To apply Eq.~\eqref{ema1} to the case in which the conducting particles have cylindrical shape, 
we denote with  $\mathbf{r}_{ij}=\mathbf{r}_i-\mathbf{r}_j$ the distance vector between the centers of mass of 
rods $i$ and $j$, and with $\mathbf{u}_i$ and $\mathbf{u}_j$ the unit vectors pointing along the axes of $i$ and $j$, 
respectively. From Eqs.~\eqref{tun6} and \eqref{iapp} we see that the tunneling conductance $g_{ij}$  
can be expressed as $g(\mathbf{r}_{ij};\mathbf{u}_i,\mathbf{u}_j)$, where the dependence on the rod
orientations is through the angle $\gamma_{ij}=\gamma(\mathbf{u}_i,\mathbf{u}_j)$ between the directions
of $\mathbf{u}_i$ and $\mathbf{u}_j$. Next, we multiply each term of the summation over $i,j$ appearing in Eq.~\eqref{ema1}
by $\int d\mathbf{u}_1d\mathbf{u}_2d\mathbf{r}_{12}
\delta(\mathbf{r}_{12}-\mathbf{r}_{ij})\delta(\mathbf{u}_1-\mathbf{u}_i)\delta(\mathbf{u}_2-\mathbf{u}_j)=1$:
\begin{align}
\label{ema2}
&\frac{1}{N}\left\langle\sum_{i,j}{}'\!\int \!d\mathbf{u}_1d\mathbf{u}_2d\mathbf{r}_{12}
\delta(\mathbf{r}_{12}-\mathbf{r}_{ij})\delta(\mathbf{u}_1-\mathbf{u}_i)\delta(\mathbf{u}_2-\mathbf{u}_j)\right.\nonumber \\
&\times\left.\frac{g(\mathbf{r}_{ij};\mathbf{u}_i,\mathbf{u}_j)}
{g(\mathbf{r}_{ij};\mathbf{u}_i,\mathbf{u}_j)+\bar{G}}\right\rangle=2,
\end{align}
and introduce the pair distribution function $P(\mathbf{r}_{12};\mathbf{u}_1,\mathbf{u}_2)$ defined as \cite{HansenMcDonald}:
\begin{align}
\label{pdf}
&\frac{\rho}{(4\pi)^2} P(\mathbf{r}_{12};\mathbf{u}_1,\mathbf{u}_2) \nonumber \\
&=\frac{1}{N}\left\langle\sum_{i,j}{}'\delta(\mathbf{r}_{12}-\mathbf{r}_{ij})
\delta(\mathbf{u}_1-\mathbf{u}_i)\delta(\mathbf{u}_2-\mathbf{u}_j)\right\rangle,
\end{align}
where $\rho$ is the number density of the cylinders. From Eqs.~\eqref{ema2} and \eqref{pdf} we 
thus rewrite the EMA equation \eqref{ema1} as follows:
\begin{equation}
\label{ema3}
\frac{\rho}{(4\pi)^2}\!\int\! d\mathbf{u}_1d\mathbf{u}_2d\mathbf{r}_{12}P(\mathbf{r}_{12};\mathbf{u}_1,\mathbf{u}_2)
\frac{g(\mathbf{r}_{12};\mathbf{u}_1,\mathbf{u}_2)}
{g(\mathbf{r}_{12};\mathbf{u}_1,\mathbf{u}_2)+\bar{G}}=2,
\end{equation}
where from Eqs.~\eqref{tun6} and \eqref{iapp}:
\begin{align}
\label{gij}
g(\mathbf{r}_{12};\mathbf{u}_1,\mathbf{u}_2)&
\equiv g\!\left[\vert\Delta\vert,z_{12};\gamma(\mathbf{u}_1,\mathbf{u}_2)\right]\nonumber \\
&=g_0 \frac{\displaystyle(L-\vert z_{12}\vert)^2e^{-2(\vert\Delta\vert-D)/\xi}}
{\displaystyle2\pi D\xi+(L-\vert z_{12}\vert)^2\sin^2\!\gamma(\mathbf{u}_1,\mathbf{u}_2)},
\end{align}
where $\vert\Delta\vert$ is the distance between the rod axes and $z_{12}=z_1-z_2$, with $\vert z_1\vert \leq  L/2$
and $\vert z_2\vert \leq  L/2$.
From Eqs.~\eqref{ema3} and \eqref{gij} we can evaluate the EMA conductance $\bar{G}$ for a given form of the
pair distribution function $P(\mathbf{r}_{12};\mathbf{u}_1,\mathbf{u}_2)$.

\section{Effect of orientational alignment}
\label{ani}

We start by considering Eq.~\eqref{ema3} for a system of randomly dispersed impenetrable rods 
with different degrees of uniaxial orientational order. To this end, we introduce
an orientational distribution function $f(\mathbf{u})$, normalized as $(4\pi)^{-1}\!\int d\mathbf{u}f(\mathbf{u})=1$,
and take the pair distribution function to have the form:
\begin{equation}
\label{pdf2}
P(\mathbf{r}_{12};\mathbf{u}_1,\mathbf{u}_2)\sim f(\mathbf{u}_1)f(\mathbf{u}_2)\theta(\vert\Delta\vert-D),
\end{equation}
where the $\theta$-function forbids the cores of the cylinders to penetrate each other, while for distances
larger than $D$ Eq.~\eqref{pdf2} assumes that the rods are completely uncorrelated. 
Since Eqs.~\eqref{gij} and \eqref{pdf2} involve $\Delta$, $z_1$, and $z_2$, it is convenient to express the integration 
over $\mathbf{r}_{12}$ in Eq.~\eqref{ema3} in terms of these variables. Using the reference frame 
$Oxyz$ defined in Fig.~\ref{fig1}, we express the vector distance between the centers of mass as
\begin{equation}
\label{r12}
\mathbf{r}_{12}=\mathbf{r}_2-\mathbf{r}_1=\left(
\begin{array}{c}
\Delta \\
-z_2\sin\!\gamma \\
z_2\cos\!\gamma-z_1
\end{array}\right),
\end{equation}
from which we get $d\mathbf{r}_{12}=\vert\sin\gamma\vert d\Delta dz_1dz_2$. Introducing the distribution function
for the angle $\gamma$:
\begin{equation}
\label{gamma1}
F(\gamma)=\int\! \frac{d\mathbf{u}_1}{4\pi}\frac{d\mathbf{u}_2}{4\pi} f(\mathbf{u}_1)f(\mathbf{u}_2)\delta[\gamma-\gamma(\mathbf{u}_1,\mathbf{u}_2)],
\end{equation}
and using Eq.~\eqref{pdf2}, the EMA equation \eqref{ema3} becomes:
\begin{align}
\label{ema4}
& 2\rho\int\! d\gamma \sin\!\gamma\, F(\gamma)\int_{-L/2}^{L/2}dz_1
\int_{-L/2}^{L/2}dz_2\int_D^\infty d\Delta\nonumber\\
&\times\frac{g(\Delta,z_{12};\gamma)}
{g(\Delta,z_{12};\gamma)+\bar{G}}=2.
\end{align}
Reducing the double integration over $z_1$ and $z_2$ to an integration over  $z=L-\vert z_1-z_2\vert$ and defining
 the dimensionless EMA conductance $g^*=\bar{G}/g_0$, Eq.~\eqref{ema4} reduces to:
\begin{align}
\label{ema5}
& 4\rho\!\left\langle\int_0^L\!\! dz\! \int_0^\infty \!\!d\delta\frac{\sin\!\gamma \,z^3e^{-2\delta/\xi}}
{(2\pi D\xi+z^2\sin^2\!\gamma)g^*+z^2e^{-2\delta/\xi}}\right\rangle_\gamma=2,
\end{align}
where $\delta=\vert \Delta\vert -D$ is the distance between the cylinder surfaces
and $\left\langle\cdots\right\rangle_\gamma=\int\!d\gamma F(\gamma)(\cdots)$
is the average over the angle $\gamma$.
We solve analytically the integrals over $\delta$ and $z$ to find:
\begin{align}
\label{ema6}
&\frac{4}{\pi}\frac{\xi L}{D^2}\phi\left\langle\sin\!\gamma
\left(1+\frac{ag^*}{1+g^*\sin^2\!\gamma}\right)
\ln\!\left[1+\frac{1}{g^*(a+\sin^2\!\gamma)}\right]\right.\nonumber \\
&\left.-\frac{a}{(1+g^*\sin^2\!\gamma)\sin\!\gamma}
\ln\!\left(\frac{a+\sin^2\!\gamma}{a}\right)\right\rangle_\gamma=2,
\end{align}
where $\phi=\rho(\pi/4)D^2L$ is the volume fraction occupied by the cylinders
and $a=2\pi D\xi/L^2$. 

To quantify the effect of tunneling anisotropy, in the following we shall compare the 
solution of Eq.~\eqref{ema6} with the EMA conductance $\bar{G}_0$ obtained 
from Eq.~\eqref{ema4} by replacing $g(\Delta,z_{12};\gamma)$ with
\begin{equation}
\label{g0}
g(\delta)=g_0e^{-2\delta/\xi},
\end{equation}
which corresponds to the tunneling conductance used in previous works \cite{Ambrosetti2010a,Nigro2013c} in which the 
tunneling processes are assumed to depend only on the relative distance between
the particles, independently of their relative orientations. Using $g(\delta)$ in Eq.~\eqref{ema4}
we find:
\begin{equation}
\label{g01}
g_0^*=\frac{\exp\!\left( -\frac{\pi}{2}\frac{D^2}{\xi L\phi\langle\sin\!\gamma\rangle_\gamma}\right)}
{1-\exp\!\left( -\frac{\pi}{2}\frac{D^2}{\xi L\phi\langle\sin\!\gamma\rangle_\gamma}\right)},
\end{equation}
where $g_0^*=\bar{G}_0/g_0$. The factor $\langle\sin\!\gamma\rangle_\gamma$ appearing in
Eq.~\eqref{g01} stems from the increased mean inter-particle distance as orientational anisotropy
is enhanced. Indeed, applying the critical path approximation \cite{cpa} to dispersions
of rods connected through $g(\delta)$, it can be shown that the system conductance for small $\phi$ is
dominated by $g(\delta_c)=g_0\exp(-2\delta_c/\xi)$, where $\delta_c$ is identified as the 
smallest distance such that the network formed by rods with $\delta\leq \delta_c$ still
spans the entire sample. Excluded volume arguments applied to systems of slender hard rods
with penetrable shells of thickness $\delta_c/2$ give
$2\delta_c/\xi=\frac{\pi}{4}\frac{D^2}{\xi L \phi\langle\sin\!\gamma\rangle_\gamma}$ \cite{Ambrosetti2010a},
which, besides a factor $2$, reproduces the argument in the exponents of Eq.~\eqref{g01}.

\begin{figure}[t]
\begin{center}
\includegraphics[scale=0.58,clip=true]{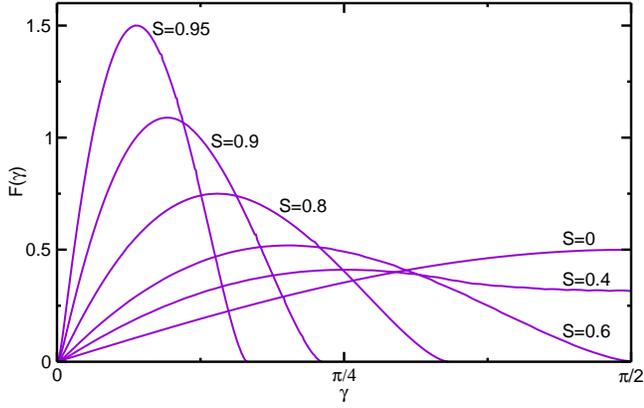}
\caption{(Color online) Distribution function $F(\gamma)$ for the angle $\gamma$
between the axes of two rods. $F(\gamma)$ is calculated numerically from Eq.~\eqref{gamma2}
using Eq.~\eqref{fu} for different values of the nematic parameter $S$. Note that 
Eq.~\eqref{fu} implies $F(\gamma)=F(\pi-\gamma)$.}\label{fig3}
\end{center}
\end{figure}

To make further progress, we consider the following model for the distribution function
of $\mathbf{u}$:
\begin{equation}
\label{fu}
f(\mathbf{u})=f(\vartheta)=\frac{\theta(\vartheta_0-\vartheta)+\theta(\vartheta_0-\pi+\vartheta)}
{1-\cos\vartheta_0},
\end{equation}
which has uniaxial symmetry $f(\vartheta)=f(\pi-\vartheta)$ and is normalized such that 
$(4\pi)^{-1}\!\int_0^{2\pi}\!d\varphi\int_0^\pi\! d\vartheta\sin\vartheta f(\vartheta)=1$. In Eq,~\eqref{fu},
$0\leq \vartheta_0\leq \pi/2$ is a cut-off angle that defines the extent of orientational order:
for $\vartheta_0=\pi/2$ the rods are oriented isotropically, while for $\vartheta_0=0$  the rods are perfectly 
aligned. It is convenient to express $\vartheta_0$ in terms of the nematic order parameter 
$S=(3\langle\cos^2\!\vartheta\rangle_\vartheta-1)/2$. From Eq.~\eqref{fu} we find:
\begin{equation}
\label{S}
S=\frac{1}{2}\cos\vartheta_0(1+\cos\vartheta_0),
\end{equation}
which varies from $S=0$ for isotropic rods to $S=1$ for perfectly aligned rods. 
Using $\cos[\gamma(\mathbf{u}_1,\mathbf{u}_2)]=\cos\vartheta_1\cos\vartheta_2+
\sin\vartheta_1\sin\vartheta_2\cos(\varphi_1-\varphi_2)$,
 where $\vartheta_i$ and $\varphi_i$ are
polar and azimuthal angles of $\mathbf{u}_i$ ($i=1,2$), we rewrite Eq.~\eqref{gamma1} as follows:
\begin{align}
\label{gamma2}
F(\gamma)=&\frac{\sin\!\gamma}{4\pi}\int_0^\pi\!d\vartheta_1\sin\vartheta_1f(\vartheta_1)
\int_0^\pi\!d\vartheta_2\sin\vartheta_2f(\vartheta_2)\nonumber\\
&\times\frac{\theta[\cos(\vartheta_1-\vartheta_2)-\cos\gamma]\theta[\cos\gamma-\cos(\vartheta_1+\vartheta_2)]}
{\sqrt{\cos(\vartheta_1-\vartheta_2)-\cos\gamma}\sqrt{\cos\gamma-\cos(\vartheta_1+\vartheta_2)}},
\end{align}
which reduces to $F(\gamma)=\sin\!\gamma/2$ for $S=0$, while numerical calculation of the double integral
in Eq.~\eqref{gamma2} reveals that $F(\gamma)$ becomes increasingly peaked as $S$ increases, as shown in 
Fig.~\ref{fig3}. Eventually, for $S=1$ the distribution function develops two Dirac-$\delta$ peaks centered at $\gamma=0$
and $\gamma=\pi$.

\begin{figure}[t]
\begin{center}
\includegraphics[scale=0.67,clip=true]{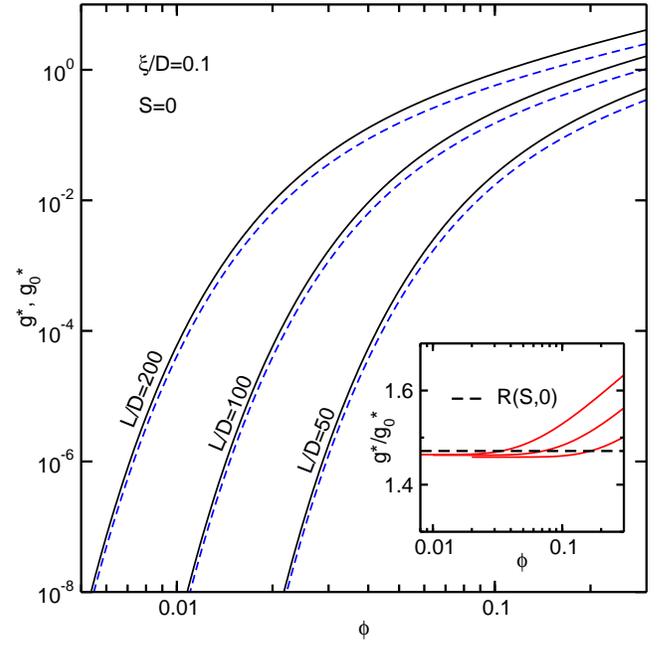}
\caption{(Color online) EMA conductance $g^*$ (solid lines) calculated from Eq.~\eqref{ema6} for the case
of isotropic rod orientations ($S=0$) and different values of $L/D$. The tunneling decay length is $\xi=0.1 D$
for all cases. Corresponding results for $g_0^*$ [Eq.~\eqref{g01}] are shown by dashed lines.
Inset: $g^*/g_0^*$ (solid lines) for $L/D=200$, $100$, and $50$, from uppermost to lowermost, compared
to $R(S=0,0)=4/e\simeq 1.47$ (dashed line)}\label{fig4}
\end{center}
\end{figure}

\begin{figure}[t]
\begin{center}
\includegraphics[scale=0.67,clip=true]{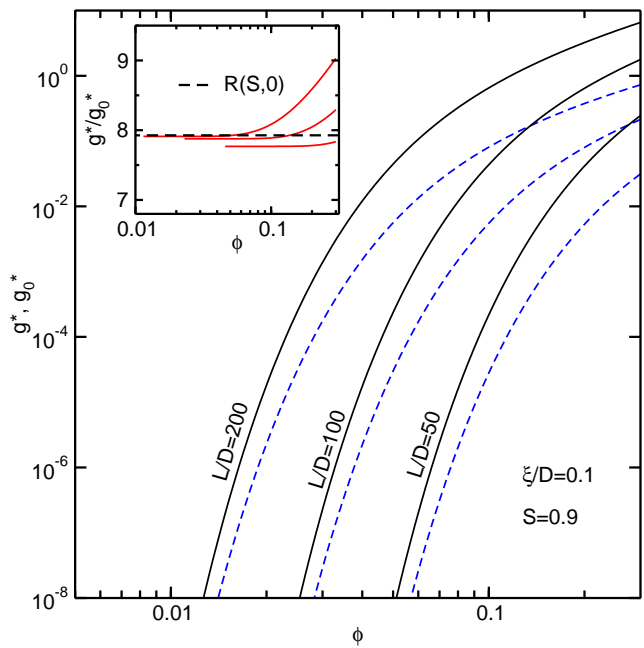}
\caption{(Color online) EMA conductance $g^*$ (solid lines) calculated from Eq.~\eqref{ema6} for strongly aligned 
rods with nematic parameter $S=0.9$ and for different values of $L/D$. The tunneling decay length is $\xi=0.1 D$
for all cases. Corresponding results for $g_0^*$ [Eq.~\eqref{g01}] are shown by dashed lines.
Inset: $g^*/g_0^*$ (solid lines) for $L/D=200$, $100$, and $50$, from uppermost to lowermost, compared
to $R(S=0.9,0)\simeq 7.93$ (dashed line) obtained from numerical calculation of Eq.~\eqref{R}}
\label{fig5}
\end{center}
\end{figure}

We solve Eq.~\eqref{ema6} numerically to find $g^*$ for different degrees
of the orientational alignment by using $F(\gamma)$ as defined above.
For isotropic orientations of the rods ($S=0$) we find that $g^*$ is only slightly enhanced 
with respect to the EMA conductance $g_0^*$ given in Eq.~\eqref{g01}, as shown in
Fig.~\ref{fig4} where we plot $g^*$ (solid lines) and $g_0^*$ (dashed lines) 
for $\xi/D=0.1$ and different $L/D$ values. From the inset of Fig.~\ref{fig4} we see that
$g^*$ is enhanced by a factor of only $1.5$-$1.6$ compared to $g_0^*$,
indicating that the role of tunneling anisotropy is marginal for isotropic orientations of the rods
even for large values of $L/D$. 
In contrast, for highly aligned rods $g^*$ is significantly enhanced compared to $g_0^*$, as 
shown in Fig.~\ref{fig5} where $g^*$ (solid lines) and $g_0^*$ (dashed lines) are plotted for 
$S=0.9$. In this case, $g^*/g_0^*$ is about $8$ or larger, as seen in the inset of Fig.~\ref{fig5}.
Furthermore, we see from Figs.~\ref{fig4} and \ref{fig5} that although
$g^*$ for $S=1$ is strongly reduced compared to the $S=0$ case for $\phi$ small, for larger 
volume fractions $g^*$ is barely affected, if not slightly enhanced, by the degree of 
orientational anisotropy. On the contrary, $g_0^*$ is reduced for all $\phi$ values as $S$ increases
from zero to $1$. 

\begin{figure}[t]
\begin{center}
\includegraphics[scale=0.67,clip=true]{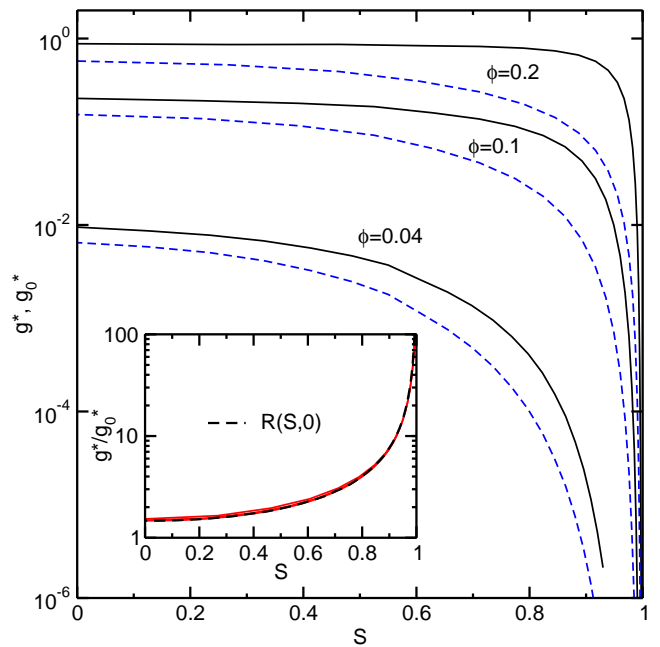}
\caption{(Color online) EMA conductances $g^*$ (solid lines) and $g_0^*$ (dashed lines) 
calculated from Eqs.~\eqref{ema6} and \eqref{g01} as functions of the nematic parameter 
$S$ and for different volume fractions. $L/D=100$ and $\xi/D=0.1$, 
for all cases. Inset: $g^*/g_0^*$ (solid lines) for the same $\phi$ values shown in the main panel,
compared to $R(S,0)$ (dashed line).}\label{fig6}
\end{center}
\end{figure}

The overall effect of the nematic order on the EMA conductance is illustrated 
in Fig.~\ref{fig6} where we show $g^*$ and $g_0^*$ as a function of $S$ for $L/D=100$,
$\xi/D=0.1$, and for different values of the volume fraction $\phi$. As $S\rightarrow 1$, both
$g^*$ (solid lines) and $g_0^*$ (dashed lines) tend to vanish, although the drop of $g_0^*$
is much faster than that of $g^*$, as evidenced by the strong increase of $g^*/g_0^*$ with $S$
shown in the inset of Fig.~\ref{fig6}. Interestingly, the data of $g^*/g_0^*$ for the same values of $\phi$
shown in the main panel of Fig.~\ref{fig6} fall approximately into a single curve, which indicates
that the net effect of the distribution of rod orientations on the tunneling anisotropy is
practically independent of the rod concentration.

To understand the behaviors shown in Figs.~\ref{fig4}-\ref{fig6} and, in particular,
the differences between $g^*$ and $g_0^*$ as $S$ is varied, we consider Eq.~\eqref{ema6} 
for very small $a$, which is the relevant limit for cylinders with high aspect-ratios:
\begin{equation}
\label{ema7}
\frac{4}{\pi}\frac{\xi L}{D^2}\phi\left\langle\sin\!\gamma
\ln\!\left[1+\frac{1}{g^*(a+\sin^2\!\gamma)}\right]\right\rangle_\gamma=2.
\end{equation}
The solution of \eqref{ema7} for $\phi\ll 1$ is:
\begin{equation}
\label{ema8}
g^*\simeq R(S,a)\exp\left(-\frac{\pi}{2}\frac{D^2}{\xi L\phi\langle\sin\!\gamma\rangle_\gamma}\right),
\end{equation}
where 
\begin{equation}
\label{R}
R(S,a)=\exp\left(-\frac{\langle\sin\!\gamma\ln (a+\sin^2\!\gamma)\rangle_\gamma}
{\langle\sin\!\gamma\rangle_\gamma}\right).
\end{equation}
Since the second term in the right-hand side of Eq.~\eqref{ema8} coincides with the dilute limit of
Eq.~\eqref{g01}, we obtain that $g^*/g_0^*\simeq R(S,a)$, at least for small $\phi$. 
$R(S,a)$ gives thus a measure of the correlation between nematic order and tunneling
anisotropy. For slender rods ($a\ll 1$) and unless
the rods are perfectly aligned \cite{notealign}, we can set $a=0$ in 
Eq.~\eqref{R}. For $S=0$, we find exactly $R(0,0)=4/e\simeq 1.47$,
where $e$ is the Neper number, which reproduces approximately the $g^*/g_0^*$ results
shown in the inset of Fig.~\ref{fig4}. For $S\neq 0$, we calculate numerically the angle averages
in Eq.~\eqref{R} to find that $R(S,0)$ increases monotonically as $S$ increases, as shown in the inset
of Fig.~\ref{fig6} (dashed line). In the same inset, we also see that $R(S,0)$ reproduces fairly 
well the quasi-universal behavior of $g^*/g_0^*$.
Clearly, $R(S,0)$ increases with $S$ because progressive alignment of the 
rods promotes enhanced tunneling processes, as illustrated in Fig.~\ref{fig2}(a). 
The enhancement of $g^*$ due to $R(S,0)$ is however opposed by the simultaneous reduction of $g_0^*$,
because $\langle\sin\!\gamma\rangle_\gamma$ diminishes as $S$ increases and eventually vanishes
at $S=1$, which explains the trend shown in Fig.~\ref{fig6}.

\section{Effect of attraction between the rods}
\label{attraction}

It is well known that in suspensions of colloidal particles the fillers may experience  
attractive forces due to the van der Waals interaction or depletion interactions 
induced by the addition of non-adsorbing polymers or surfactant micelles \cite{Lekkerkerker2011}.
When the fillers are conducting, attractive forces may change drastically the conductivity
of the composite as compared to that of fluids of hard-core particles, as attraction
promotes enhanced tunneling between the fillers. For spherical conducting
colloidal particles with square-well attractive potentials, numerical simulations have
evidenced increased tunneling conductivity both for equilibrium fluids \cite{Nigro2012a} and 
kinetically arrested gels \cite{Nigro2013b}. Experimentally, enhanced electrical connectivity
due to depletion interaction has been observed in polymer composites with carbon nanotubes \cite{Vigolo2005,Maillaud2013}
and in silver/epoxy nanocomposites with added silica particles \cite{Cho2014}.

When the shape of the colloidal particles is rod-like, effectively short-range attractive forces
become highly anisotropic because, at separations smaller than the attraction range, two parallel 
rods have larger surface of interaction compared to skewed ones \cite{Schoot1992}. 
Short-range attraction promotes thus parallel configurations.
Although the functional form of the interaction as a function of 
the relative rod distance depends on the specific mechanism of attraction \cite{Schoot1992,Odijk1994,Sear1997}, the
dependence on the mutual orientation between the rods is rather generic and scales as
$\sin^{-1}\!\gamma_{ij}$ for two cylinders $i$ and $j$ skewed by an angle 
$\gamma_{ij}=\gamma(\mathbf{u}_i,\mathbf{u}_j)$ \cite{Kyrylyuk2008,Schoot1992,Odijk1994,Lekkerkerker1997,Schiller2010,Sear1997}. 
For short-ranged attractions between impenetrable and perfectly rigid cylinders, we adopt 
here a square-well potential $W_{ij}=W(\mathbf{r}_{ij};\mathbf{u}_i,\mathbf{u}_j)$ 
which has been previously used to describe depletion interaction \cite{Kyrylyuk2008,Schoot1992}.
We thus take $W_{ij}$ to have hard-core repulsion $W_{ij}=\infty$ for $\vert\Delta_{ij}\vert<D$,
attraction $W_{ij}=W^\textrm{A}(L-\vert z_{ij}\vert,\gamma_{ij})$ 
for $D\leq \vert\Delta_{ij}\vert\leq D(1+\lambda)$, and $W_{ij}=0$ for 
$\vert\Delta_{ij}\vert>D(1+\lambda)$, where $\vert\Delta_{ij}\vert$ is the distance between the 
rod axes as given in Fig.~\ref{fig1}, $\lambda D\ll D$ is the range of attraction, and 
\begin{equation}
\label{W1}
\beta W^\textrm{A}(L-\vert z_{ij}\vert,\gamma_{ij})\sim\left\{
\begin{array}{cc}
\displaystyle-\frac{\sqrt{\lambda}\varepsilon}{D}(L-\vert z_{ij}\vert), & \gamma_{ij}< \frac{\sqrt{\lambda}D}{L} \\
\displaystyle-\frac{\lambda\varepsilon}{\sin\!\gamma_{ij}}, & \gamma_{ij}> \frac{\sqrt{\lambda}D}{L}
\end{array}\right.
\end{equation}
is the attraction well, where $z_{ij}=z_i-z_j$, $\beta$ is the inverse temperature, 
and $\varepsilon\geq 0$ is the  dimensionless strength of the attraction. 
We immediately see from Eq.~\eqref{W1} and from
Eqs.~\eqref{i2} and \eqref{i3} that  $W^\textrm{A}_{ij}$ and the tunneling matrix element 
$\vert I_{ij}\vert$ have strikingly similar dependencies on $\gamma_{ij}$ and on 
$\vert z_i-z_j\vert$. This correspondence is not totally unexpected, since both $W^\textrm{A}_{ij}$
and $\vert I_{ij}\vert$ are proportional to areas of overlap: for the case of tunneling the overlap is
between the wave functions of the two rods, while for the attraction the overlap is given by the 
potential range. From this observation, we infer thus that tunneling between two rods within 
the attraction range is enhanced, since $W^\textrm{A}_{ij}$ promotes alignment of the rods.

To find the EMA conductance for dispersions of attractive rods, we must specify the
pair distribution function $P(\mathbf{r}_{12};\mathbf{u}_1,\mathbf{u}_2)$ appearing in 
Eq.~\eqref{ema3}. It is well known that equilibrium distributions of attractive rods display
different phases depending on the rod concentration, strength of interaction, and 
$L/D$ \cite{Kyrylyuk2008,Lekkerkerker2011,Lekkerkerker1997,Melgar2009}.
In fluids with sufficiently small $\phi$ and weak attractions, rods have isotropic orientations and
local correlations. In this regime we approximate the pair distribution function by its
low-density limit \cite{HansenMcDonald}: 
\begin{equation}
\label{PA}
P(\mathbf{r}_{12};\mathbf{u}_1,\mathbf{u}_2)\simeq e^{\displaystyle-\beta W(\mathbf{r}_{12};\mathbf{u}_1,\mathbf{u}_2)},
\end{equation}
which reduces Eq.~\eqref{ema3} to:
\begin{equation}
\label{ema9}
\frac{\rho}{(4\pi)^2}\!\int\! d\mathbf{u}_1d\mathbf{u}_2d\mathbf{r}_{12}
\frac{e^{\displaystyle-\beta W(\mathbf{r}_{12};\mathbf{u}_1,\mathbf{u}_2)}
g(\mathbf{r}_{12};\mathbf{u}_1,\mathbf{u}_2)}
{g(\mathbf{r}_{12};\mathbf{u}_1,\mathbf{u}_2)+\bar{G}}=2.
\end{equation}
The dependence of the attraction potential on $\Delta$, $z_{12}$, and $\gamma$ allows us 
to follow the same steps outlined in Sec.~\ref{ani}. We thus express $\mathbf{r}_{12}$ in 
terms of $(\Delta,z_1,z_2)$, reduce the double integral over $z_1$ and $z_2$ to an
integral over $z=L-\vert z_1-z_2\vert$, and integrate over the distance $\Delta$ to find:
\begin{widetext}
\begin{align}
\label{ema10}
&\frac{8\phi\xi}{\pi D^2L}\left\langle\sin\!\gamma\int_0^L\!dz\, z e^{\displaystyle-\beta W^\textrm{A}(z,\gamma)}
\ln\!\left[\frac{z^2+(z^2\sin^2\!\gamma+2\pi D\xi)g^*}{z^2\exp(-2D\lambda/\xi)+(z^2\sin^2\!\gamma+2\pi D\xi)g^*}\right]\right\rangle_\gamma\nonumber \\
&+\frac{8\phi\xi}{\pi D^2L}\left\langle\sin\!\gamma\int_0^L\!dz\, z 
\ln\!\left[\frac{z^2\exp(-2D\lambda/\xi)+(z^2\sin^2\!\gamma+2\pi D\xi)g^*}{(z^2\sin^2\!\gamma+2\pi D\xi)g^*}\right]\right\rangle_\gamma=2,
\end{align}
\end{widetext}
where $g^*=\bar{G}/g_0$ and the average over $\gamma$ is done
over an isotropic distribution of the rod orientations, i.e., $F(\gamma)=\sin\!\gamma/2$. 
From Eq.~\eqref{W1} we see that $\exp[-\beta W^\textrm{A}(z,\gamma)]\simeq 1$ 
for shallow well potentials such that $\sqrt{\lambda}\varepsilon L/D\lesssim 1$, and
Eq.~\eqref{ema10} reduces basically to the $S=0$ case studied in Sec.~\ref{ani}. 
The interesting situation arises when $\sqrt{\lambda}\varepsilon L/D\gg 1$ which makes
 $\exp[-\beta W^\textrm{A}(z,\gamma)]$ to basically select only parallel rod configurations \cite{Schoot1992}.
In this case, we estimate the dominant contribution to the first term in Eq.~\eqref{ema10}
by setting $\gamma=0$ and $z=L$ in the argument of the logarithm.
The integration over $z$ in the second term of Eq.~\eqref{ema10} can be done exactly and for small
$a=2\pi D\xi/L^2$ the result coincides
with the left-hand side of Eq.~\eqref{ema7} with $g^*$ replaced by $g^*\exp(2D\lambda/\xi)$.
We thus find:
\begin{align}
\label{ema11}
&\frac{\xi L}{D^2}\phi\frac{\chi+\lambda}{\lambda}\ln\!\left(\frac{1+ag^*}{e^{-2D\lambda/\xi}+ag^*}\right)\nonumber \\
&+\frac{4}{\pi}\frac{\xi L}{D^2}\phi\left\langle \sin\!\gamma
\ln\!\left[1+\frac{e^{-2D\lambda/\xi}}{g^*(a+\sin^2\!\gamma)}\right]\right\rangle_\gamma=2,
\end{align}
where we have introduced the variable $\chi=-B^\textrm{A}/B^\textrm{hc}$, in which
$B^\textrm{hc}=\pi DL^2/4$ and
\begin{equation}
\label{chi}
B^\textrm{A}=-2D\lambda\left\langle\sin\!\gamma\int_0^L\! dz\,z 
\left[e^{\displaystyle-\beta W^\textrm{A}(z,\gamma)}-1\right]\right\rangle_\gamma
\end{equation}
are respectively the contributions of the hard-core
and of the attraction well to the second-virial coefficient $B=B^\textrm{hc}+B^\textrm{A}=-\frac{1}{2}\int d\mathbf{u}_1d\mathbf{u}_2
d\mathbf{r}_{12}[e^{-\beta W(\mathbf{r}_{12};\mathbf{u}_1,\mathbf{u}_2)}-1]$ \cite{Schoot1992,HansenMcDonald}. 
From Eq.~\eqref{ema11} we see thus that, for a given range of the potential, $g^*$ does not
depend on the details of the attraction well, at least as long as $\sqrt{\lambda}\varepsilon L/D\gg 1$.

We find that the solution of Eq.~\eqref{ema11} for $g^*\lesssim a^{-1}\exp(-2\lambda D/\xi)$ reduces to:
\begin{equation}
\label{ema12}
g^*\simeq R(0,a)\exp\left(\frac{2D\chi}{\xi}\right) \exp\left(-\frac{2D^2}{\phi\xi L}\right),
\end{equation}
where $R(0,a)\simeq 4/e$ is given by the $S=0$ limit of Eq.~\eqref{R}. In terms of volume fraction, Equation \eqref{ema12}
applies when $\phi\lesssim \phi^*$, where 
\begin{equation}
\label{phistar}
\phi^*=\frac{D/L}{\chi+\lambda-(\xi/2D)\ln(e/4a)},
\end{equation}
which we obtain by equating Eq.~\eqref{ema12} to $a^{-1}\exp(-2\lambda D/\xi)$. 
For $g^*\gtrsim a^{-1}\exp(-2\lambda D/\xi)$ (i.e., for $\phi\gtrsim\phi^*$) and for $\chi/\lambda\gtrsim 1$ we neglect the 
second term in the left-hand side of Eq.~\eqref{ema11} to find:
\begin{equation}
\label{ema13}
g^*\simeq \frac{L^2}{2\pi\xi D}\frac{\exp\!\left(-\frac{2D^2}{\phi\xi L}\frac{\lambda}{\chi+\lambda}\right)-\exp\!\left(-\frac{2D\lambda}{\xi}\right)}
{1-\exp\!\left(-\frac{2D^2}{\phi\xi L}\frac{\lambda}{\chi+\lambda}\right)}.
\end{equation}

To illustrate the net effect of the tunneling anisotropy, we compare the two limiting behaviors of Eqs.~\eqref{ema12} and 
\eqref{ema13} with those arising by considering a tunneling conductance which depends only on the relative distance
between two cylinders, as done in Sec.~\ref{ani}. Using Eq.~\eqref{g0} in Eq.~\eqref{ema9}, and following 
the same steps outlined above to solve the integrals, we find that the resulting dimensionless EMA conductance 
$g_0^*$ satisfies the following exact relation: 
\begin{equation}
\label{ema14}
\frac{\phi\xi L}{D^2}\left[\frac{\chi}{\lambda}\ln\!\left(\frac{1+g_0^*}{e^{-2\lambda D/\xi}+g_0^*}\right)
+\ln\!\left(\frac{1+g_0^*}{g_0^*}\right)\right]=2,
\end{equation}
which gives either:
\begin{equation}
\label{ema15}
g_0^*\simeq \exp\left(\frac{2D\chi}{\xi}\right) \exp\left(-\frac{2D^2}{\phi\xi L}\right),
\end{equation}
for $g_0^*\lesssim \exp(-2D\lambda/\xi)$, or:
\begin{equation}
\label{ema16}
g_0^*\simeq \frac{\exp\!\left(-\frac{2D^2}{\phi\xi L}\frac{\lambda}{\chi+\lambda}\right)}
{1-\exp\!\left(-\frac{2D^2}{\phi\xi L}\frac{\lambda}{\chi+\lambda}\right)},
\end{equation}
for $g_0^*\gtrsim \exp(-2D\lambda/\xi)$. As a function of volume fraction, Equations \eqref{ema15}
and \eqref{ema16} apply either when $\phi\lesssim \phi_0^*$ or $\phi\gtrsim\phi_0^*$, respectively, where:
\begin{equation}
\label{phi0star}
\phi_0^*=\frac{D/L}{\chi+\lambda}.
\end{equation}

\begin{figure}[t]
\begin{center}
\includegraphics[scale=0.58,clip=true]{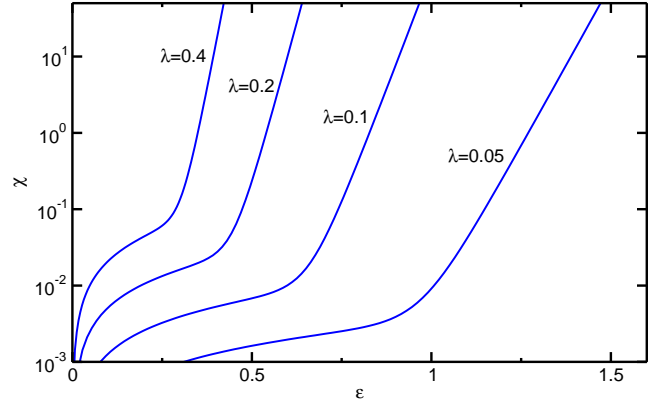}
\caption{(Color online) Reduced second-virial coefficient 
$\chi=\frac{8\lambda}{\pi L^2}\left\langle\sin\!\gamma\int_0^Ldz z\exp[-\beta W^\textrm{A}(z,\gamma)]\right\rangle_\gamma-\lambda$,
with $W^\textrm{A}(z,\gamma)$ given in Eq.~\eqref{W2}. $\chi$ is shown as a function of the potential depth $\varepsilon$
for different values of the range $\lambda$, and for aspect-ratio fixed at $L/D=100$.}
\label{fig7}
\end{center}
\end{figure}

When we compare Eq.~\eqref{ema12} with Eq.~\eqref{ema15}, and Eq.~\eqref{ema13} with Eq.~\eqref{ema16}, 
we see that $g^*$ is systematically enhanced with respect to $g_0^*$, and that this enhancement 
depends on $\phi$ according to:
\begin{equation}
\label{ema17}
\frac{g^*}{g_0^*}\simeq\left\{
\begin{array}{cc}
4/e, & \phi\lesssim \phi_0^*,\\
L^2/(2\pi\xi D), & \phi\gtrsim \phi^*,
\end{array}\right.
\end{equation}
where for $\phi\gtrsim \phi^*$ we have neglected the term $\exp(-2D\lambda/\xi)$ 
appearing in the numerator of Eq.~\eqref{ema13}. The above relation means that
in systems of attracting rods the net effect of tunneling anisotropy is marginal
in the low density region $\phi\lesssim \phi_0^*$, while it becomes remarkably strong for
larger concentrations of slender rods. Interestingly, for dispersions of cylinders
with $L/D\approx 100$ and $\xi/D\approx 0.1$, equations 
\eqref{phistar} and \eqref{ema17} predict an enhancement factor of about four orders of 
magnitude for concentrations larger than only $1$-$2$ \%, even for moderate attractions
of order $\chi=\mathcal{O}(1)$.

To assess the accuracy of the approximate EMA conductances obtained above, we should
consider a more complete functional form of $W^\textrm{A}_{ij}$ than the partial one given in Eq.~\eqref{W1}
to solve numerically Eqs.~\eqref{ema10} and \eqref{ema14}. To this end, it suffices
to consider an ansatz for $W^\textrm{A}_{ij}$ which reproduces the limiting behaviors of Eq.~\eqref{W1},
as $g^*$ and $g_0^*$ do not depend on the details of the potential well, at least for
attractions peaked at small $\gamma$.  We thus take: 
\begin{equation}
\label{W2}
\beta W^\textrm{A}(L-\vert z_{ij}\vert;\gamma_{ij})
=-\frac{\lambda\varepsilon(L-\vert z_{ij}\vert)}
{\sqrt{D^2\lambda+(L-\vert z_{ij}\vert)^2\sin^2\!\gamma_{ij}}},
\end{equation}
from which we calculate numerically for different values of $\epsilon$ and $\lambda$ the reduced 
second-virial coefficient $\chi$ shown in Fig.~\ref{fig7}. On enhancing $\varepsilon$ for a given 
$\lambda$, $\chi$ crosses over an exponential behavior of the form 
$\chi\propto \exp(\sqrt{\lambda}\varepsilon L/D)$, which signals that rods within the potential range
have mainly parallel configurations \cite{Schoot1992}. It is in this regime that $g^*/g_0^*\propto L^2/(D\xi)$ 
is expected to hold true when $\phi\gtrsim\phi^*$.

\begin{figure}[t]
\begin{center}
\includegraphics[scale=0.67,clip=true]{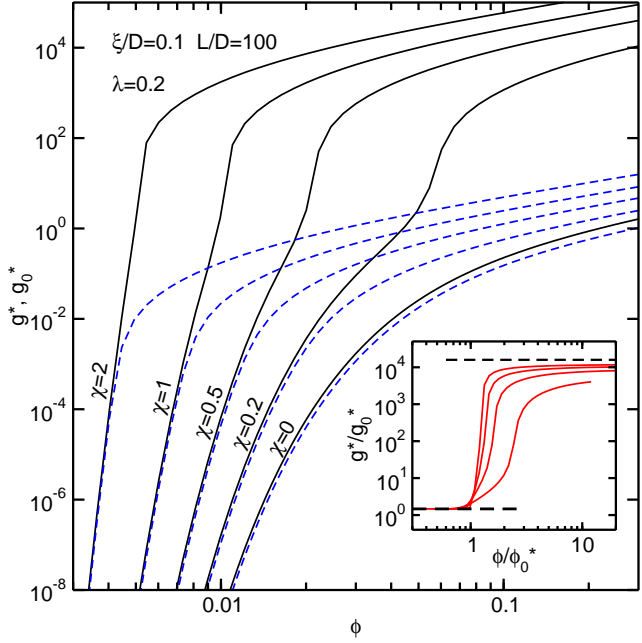}
\caption{(Color online) EMA conductances $g^*$ (solid lines) and $g_0^*$ (dashed lines)
obtained from numerical solution of Eqs.~\eqref{ema10} and \eqref{ema14}, respectively,
with the attraction well potential given in Eq.~\eqref{W2}. Different pairs of solid and dashed lines 
are calculated for different reduced second-virial coefficients $\chi$, with potential range fixed at 
$\lambda=0.2$. $L/D=100$ and $\xi/D=0.1$, for all cases. Inset: enhancement factor $g^*/g_0^*$
as a function of $\phi/\phi_0^*$, where $\phi_0^*$ is given in Eq.~\eqref{phi0star}. $\chi=2$, $1$, $0.5$,
and $0.2$ from the uppermost to the lowermost curve. Dashed horizontal lines indicate
$4/e\simeq 1.47$ and $L^2/(2\pi D\xi)\simeq 15915$.}
\label{fig8}
\end{center}
\end{figure}

\begin{figure}[t]
\begin{center}
\includegraphics[scale=0.67,clip=true]{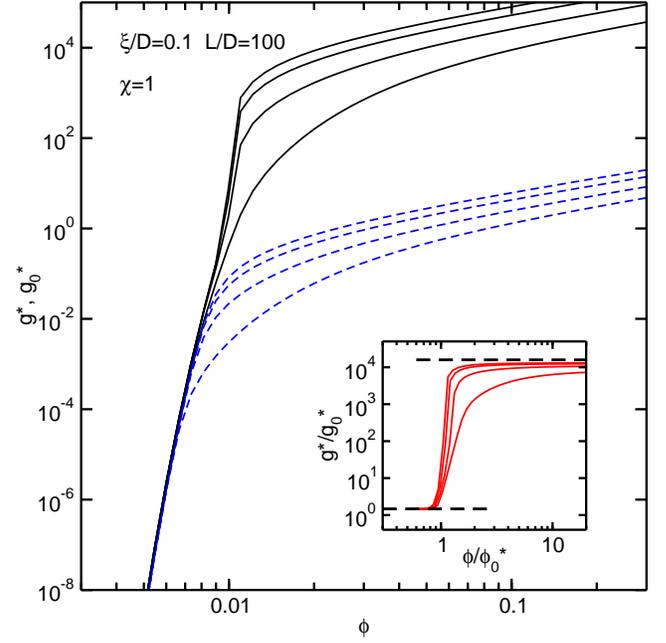}
\caption{(Color online) EMA conductances $g^*$ (solid lines) and $g_0^*$ (dashed lines)
obtained from numerical solution of Eqs.~\eqref{ema10} and \eqref{ema14}, respectively,
with the attraction well potential given in Eq.~\eqref{W2}. The attraction range is 
$\lambda=0.05$, $0.1$, $0.2$, and $0.4$ from the uppermost to the lowermost solid
and dashed lines, while the reduced second-virial coefficient is fixed at $\chi=1$.
$L/D=100$ and $\xi/D=0.1$, for all cases. Inset: enhancement factor $g^*/g_0^*$
as a function of $\phi/\phi_0^*$, where $\phi_0^*$ is given in Eq.~\eqref{phi0star}. $\lambda=0.05$, $0.1$, $0.2$,
and $0.4$ from the uppermost to the lowermost curve. Dashed horizontal lines indicate
$4/e\simeq 1.47$ and $L^2/(2\pi D\xi)\simeq 15915$.}
\label{fig9}
\end{center}
\end{figure}

Using the attraction well of Eq.~\eqref{W2}, we solve numerically Eqs.~\eqref{ema10} and \eqref{ema14} 
to calculate the EMA conductances $g^*$ and $g_0^*$ shown respectively by solid and dashed lines
in Fig.~\ref{fig8}, where the potential range is fixed at $\lambda=0.2$ and $\chi$ varies between $0$ and $2$.
For all cases, $L/D=100$ and $\xi/D=0.1$. The case $\chi=0$ in Fig.~\ref{fig8} corresponds
to the situation without attraction (i.e., $\varepsilon=0$), and the resulting $g^*$ and $g_0^*$ are the 
same as those shown in Fig.~\ref{fig3} for isotropic orientations of the rods. As $\chi$ increases for
fixed $\lambda$, both conductances are enhanced with respect to the case with no attraction.
In particular, $g^*$ and $g_0^*$ drop to low levels of conductivities at increasingly smaller volume fractions as
the reduced second-virial coefficient grows from $\chi=0$ to $\chi=2$, as shown in Fig.~\ref{fig8}.
For $\phi$ smaller than a characteristic volume fraction, which is well approximated by $\phi_0^*$
of Eq.~\eqref{phi0star}, $g^*$ closely follow $g^*\simeq (4/e) g_0^*$, as predicted by Eq.~\eqref{ema17} and
shown in the inset of Fig.~\ref{fig8}. As the volume fraction becomes lager than about $\phi_0^*$, $g^*$ increases 
much more rapidly than $g_0^*$ and eventually reaches a level of conductivity which is about four orders
of magnitude larger than $g_0^*$, as illustrated in the inset of Fig.~\ref{fig8}. In this regime, the effect
of tunneling anisotropy is largest and $g^*/g_0^*$ is proportional to $L^2/(D\xi)$, confirming the result
of Eq.~\eqref{ema17} for $\phi\gtrsim \phi^*$. As a further confirmation of the previous analysis, we note
that such strong enhancement of $g^*$ compared to $g_0^*$ is attained even for moderate values of the 
reduced second-virial coefficient: $g^*/g_0^*\gg 1$ already for
$\chi\simeq 0.5$-$1$ for volume fractions larger than $1$-$2$ \%.

For $\phi\lesssim\phi_0^*$, $g^*$ and $g_0^*$ depend
solely on the second-virial coefficient, independently of the range of attraction, as shown in Fig.~\ref{fig9}
where the EMA conductances are shown for different values of $\lambda$ and $\varepsilon$ chosen so to give $\chi=1$.
On the contrary, for larger $\phi$ values both conductances become affected by the attraction potential profile, 
in accord with the predictions of Eqs.~\eqref{ema13} and \eqref{ema16}. Similar results have been found previously for
the case of conducting spheres attracting via a square-well potential \cite{Nigro2012a}.
 
We have repeated the above analysis by considering attraction potentials different from Eq.~\eqref{W2} but that
reduce to the limits given in Eq.~\eqref{W1}. In particular, using in Eq.~\eqref{ema10}
$\beta W^\textrm{A}(z,\gamma)=-\lambda\varepsilon z/(D^n\lambda^{n/2}+z^n\sin^n\gamma)^{1/n}$ with $n$
integer and positive, we have verified that the resulting $g^*$ is practically independent of the choice for $n$, confirming thus
the essential independence of the EMA conductance on the particular form of $W^\textrm{A}$ for given
$\chi$ and $\lambda$.

The observation that the effect of tunneling anisotropy is weak as $\phi$ goes to zero, (i.e., that $g^*/g_0^*\simeq 1.47$)
has interesting consequences when we allow the rods to be dispersed within an insulating medium that has a small but
nonzero conductivity $\sigma_\textrm{ins}$. In this case, the dimensionless EMA conductance of the total system
constituted by the rods and the insulating medium is limited from below by $g_\textrm{ins}^*=\bar{G}_\textrm{ins}/g_0$, 
where $\bar{G}_\textrm{ins}$ represents the EMA equivalent of $\sigma_\textrm{ins}$ \cite{Ambrosetti2010a,Nigro2012a}. 
According to the previous analysis and to Figs.~\ref{fig8} and \ref{fig9}, for $g_\textrm{ins}^*<\exp(-2\lambda D/\xi)$ 
the location of the conductor-insulator crossover point is:
\begin{equation}
\label{phic}
\phi_c\simeq \frac{D/L}{\chi+(\xi/2D)\ln(1/g_\textrm{ins}^*)},
\end{equation}
which we obtain by equating Eq.~\eqref{ema12} or \eqref{ema15} with $g_\textrm{ins}^*$ (we neglect the
unimportant factor $4/e$). We see thus that the crossover position does not determine the behavior of 
either $g^*$ or of $g_0^*$ for volume fractions larger than $\phi_c$, as Eq.~\eqref{phic} depends on the square well potential
only through $\chi$. More importantly, even if $g^*$ and $g_0^*$ share the same $\phi_c$, they have
a completely different functional dependence for volume fractions sufficiently larger than $\phi_c$, as
clearly illustrated in Figs.~\ref{fig8} and \ref{fig9}. These considerations are particularly relevant when
we realize that the operational definition of the percolation threshold in experiments on real nanocomposites 
is given by the value of $\phi$ below which the composite conductivity matches that of the insulating phase
(or, alternatively, the lowest measurable conductivity) \cite{Ambrosetti2010a}, just as we derived Eq.~\eqref{phic}.
Measurements of the percolation threshold in composites of attractive rods are thus not expected to give 
evidence of tunneling anisotropy effects.

We have derived the above results by using the low density limit Eq.~\eqref{PA} for the pair
distribution function. However, even for small values of the reduced second-virial coefficient,
higher order terms involving three or more bodies in Eq.~\eqref{PA} cannot be neglected when $L/D\gg 1$ \cite{Schoot1992}.
Nevertheless, for isotropic liquids of attractive rods, the EMA equations \eqref{ema11} and \eqref{ema14} may 
still be used if we re-interpret $\chi$ as the normalized contact value of the pair distribution function. 
For short-ranged square-well potentials, $P(\mathbf{r}_{12},\mathbf{u}_1,\mathbf{u}_2)$ is indeed not continuous at 
the edge of the attraction potential \cite{Haya2003}, and can be expressed as $P(\mathbf{r}_{12},\mathbf{u}_1,\mathbf{u}_2)=0$
for $\Delta<D$, 
$P(\mathbf{r}_{12},\mathbf{u}_1,\mathbf{u}_2)=P^\textrm{A}(\mathbf{r}_{12},\mathbf{u}_1,\mathbf{u}_2)$
for $D\leq \Delta\leq D(1+\lambda)$, and
$P(\mathbf{r}_{12},\mathbf{u}_1,\mathbf{u}_2)=P^\textrm{out}(\mathbf{r}_{12},\mathbf{u}_1,\mathbf{u}_2)$ for $\Delta>D(1+\lambda)$
(i.e., outside the potential well).
For $\sqrt{\lambda}\varepsilon L/D\gg 1$, we expect that $P^\textrm{A}$ is strongly peaked at $\gamma\simeq 0$, 
while $P^\textrm{out}\sim 1$. 
Assuming that $P^\textrm{A}$ can be written approximately as $P^\textrm{A}(\Delta,z,\gamma)$, where 
$z=L-\vert z_1-z_2\vert$, and considering that $P^\textrm{A}(\Delta,z,\gamma)$ depends weakly on  $\Delta$ for $\lambda$ small,
and that its dominant contribution is for $z\simeq L$ and $\gamma\simeq 0$, we can still write Eqs.~\eqref{ema11}
and \eqref{ema12}, where now:
\begin{equation}
\label{chi2}
\chi\simeq \frac{8}{\pi D L^2}\left\langle\sin\!\gamma\int_0^Ldz z\int_D^{D(1+\lambda)}
\left[P^\textrm{A}(\Delta,z,\gamma)-1\right]\right\rangle_\gamma.
\end{equation}

\section{Discussion and conclusions}
\label{concl}

The tunneling anisotropy in rod-like conducting particles induced by the relative orientation of the rod axes
is a quantum mechanical effect which, to the best of our knowledge, has not been considered so far
in the study of the electron transport in nanorod systems. We have shown that the tunneling matrix element 
of parallel configurations of two slender cylindrical particles is about $L/\sqrt{D\xi}$ times
larger than the matrix element of perpendicular cylinders. This strong orientational dependence
of tunneling has interesting consequences for the conductivity of nanorod suspensions. Namely, as follows.

(i) For isotropic distributions of rod orientations, the inclusion of the angular dependence in tunneling
 induces only a marginal increase of the system conductivity compared to the conductivity $g_0^*$ 
 in which tunneling anisotropy is ignored. On the contrary, in systems with increased orientational ordering, 
 the conductivity $g^*$ with full angular dependence is significantly enhanced compared to $g_0^*$.

(ii) Tunneling anisotropy induces a strong increase of the conductivity when the rods interact
via an attractive, short-range, potential. Depending on the potential profile, the increase compared
to the case in which the angular dependence of the inter-rod conductance is neglected is proportional
to about $L^2/D\xi$ when the volume fraction is larger than $\phi^*$ given in Eq.~\eqref{phistar}.

These features illustrate that tunneling anisotropy may have remarkable effects, especially in the functional 
dependence on $\phi$ of the system conductivity $g^*$, as discussed in Sec.~\ref{attraction}, where we have shown
that the position $\phi_c$ of the ``percolation" (or, more correctly, the crossover) transition to the insulating regime 
is barely affected by the tunneling anisotropy, which instead dominates transport at larger volume fractions.
In this respect, we note that values of the maximum conductivity $\sigma_\textrm{max}$ measured in nanotube and 
nanofiber composites may vary by several orders of magnitude even for systems with similar values of $\phi_c$ 
and of aspect-ratios \cite{Bauhofer2009,Ambrosetti2010a}. In addition to changes in morphology induced by
the nature of the insulating matrix and the method of preparation, tunneling anisotropy could possibly be
a further source of the observed scatter of $\sigma_\textrm{max}$ values. 

Our results rest on a few assumptions and simplifications that we summarize as follows.

(1) The rod-like particles are modeled by slender cylinders with $L/D\gg 1$ and the tunneling decay length $\xi$
is assumed to be much smaller than the cylinder diameter $D$. These two assumptions allow us to neglect details
of the cylinder ends and to consider tunneling only between the lateral walls of the cylinders. 
Furthermore, they permit us to neglect the coupling between different states associated to the radial
wave functions, simplifying considerably the expression for the tunneling matrix element.  
Noting that for composites with polymeric matrices the typical values of $\xi$ range from a fraction of a nanometer to 
a few nanometers,  $\xi/D\ll1$ is appropriate for carbon and metallic nanofibers or for multi-walled
carbon nanotubes, as these particles have $D$ typically larger than a few tens nanometers.

(2) The cylinders are perfectly rigid and straight. Although this assumption is generally
appropriate for nanofiber and nanowires composites, it is certainly less accurate, or even insufficient, for polymer
nanocomposites filled with carbon nanotubes, as these usually display a high degree of waviness stemming from their
intrinsic bending flexibility. In the case of tunneling between two curved cylinders, the notions of parallel or perpendicular
configurations lose their meaning, and we cannot apply the tunneling matrix element analysis of Sec.~\ref{sec:matrix}.
However, we can still tentatively use the formalism here introduced when the persistence length $L_\textrm{P}$ of
flexible nanotubes is much larger than their physical length (or contour length) $L$. When $L_\textrm{P}/L\gg  1$, two 
nanotubes at the point of closest approach may be approximated by straight cylinders, and the resulting tunneling 
matrix element should thus not differ much from Eqs.~\eqref{i1}. 

(3) The lengths and diameters of all particles are identical. We note that
the connectivity of rods with length polydispersity may strongly depend on
the length distribution \cite{Otten2009,Chatterjee2010,Mutiso2012,Nigro2013a}. 
The calculation of the tunneling matrix element between two rods of lengths $L_i$ and $L_j$,
both assumed to be much larger than $D$, follows the same steps detailed in Sec.~\ref{sec:matrix}. Here,
it suffices to note that the matrix element $I_{ij}$ for two parallel rods is still given by Eq.~\eqref{i2} with $L$ replaced
by $\min(L_i,L_j)$, while $I_{ij}$ given in Eq.~\eqref{i3} for perpendicular configurations remains unaltered.

(4) Dispersions of attractive rods are homogeneous and have isotropic orientations of the rod axes. 
This approximation is appropriate for small rod concentrations and relatively weak attraction potentials. 
We speculate that, for fixed attraction, the onset of nematic order as $\phi$ increases would possibly give
a nonmonotonic behavior of the conductivity, with a maximum centered about a concentration value that depends
on $L/D$, $\lambda$, and $\varepsilon$. Similarly, a maximum of conductivity is also expected at fixed $\phi$ as 
attraction is enhanced, because stronger attractions induce nematic order or the formation of rod bundles, which
can be viewed as particles with lower aspect-ratios than that of isolated rods. This effect has been recently observed
in composites of carbon nanotubes with added surfactant micelles \cite{Maillaud2013}. 
We expect that the inclusion of tunneling anisotropy would enhance the value of the maximum conductivity without 
shifting its position. This scenario could be verified within the EMA approach by using in Eq.~\eqref{ema3} pair 
distribution functions extracted from simulations of attractive rods, in the same way as done in Ref.~\cite{Nigro2013c}.

We conclude by pointing out that tunneling anisotropy could have important effects also for anisometric particles
other than rod-like ones. In particular, conducting fillers with disk-like shapes, as graphite or graphene nanoplatelets,
may display even stronger tunneling anisotropy effects than those described in this paper. On physical grounds,
we expect indeed that tunneling between the wave functions associated to two parallel disks facing each other would 
extend over the whole overlapping \emph{area} of the disks, while for the case of parallel rods tunneling is limited to the
overlapping \emph{length} of the cylinders. In addition, electrical connectedness of equilibrium
distributions of disks competes with nematic order in a much wider range of aspect ratios compared to the case of
rod systems \cite{Mathew2012}. In this situation, parallel configurations of the disks are predominant, and we expect that
tunneling gets enhanced.  

We thank Avik P. Chatterjee for useful comments.
B. N. acknowledges support by the Swiss National Science Foundation (Grant No. 200020-135491).

\appendix
\section{Study of the tunneling matrix element and comparison with Eq.~\eqref{i1}}
\label{appa}

\begin{figure}[t]
\begin{center}
\includegraphics[scale=0.84,clip=true]{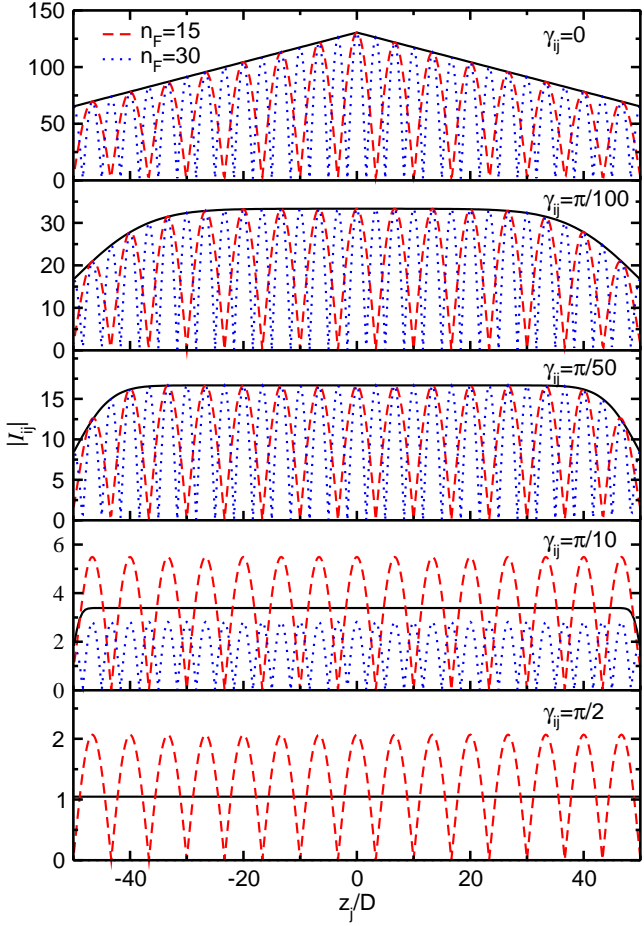}
\caption{(Color online) Dimensionless tunneling matrix element $\vert I_{ij}\vert$ of
Eq.~\eqref{i0} as a function of the misalignment $z_{ij}=z_i-z_j$ (with $z_i=0$ and $-L/2\leq z_j 
\leq L/2$) of the centers of mass of two cylinders with $L/D=100$, $\Delta_{ij}=D$, and $\xi/D=0.1$.
The Fermi wave numbers are $k_F=\pi n_F/L$ with $n_F=15$ (dashed lines) and $n_F=30$ (dotted lines).
The angle between the rod axes is $\gamma_{ij}=0$, $\pi/100$, $\pi/50$, $\pi/10$, and $\pi/2$ (panels from top to bottom).
$\vert I_{ij}\vert$ is identically zero for $\gamma_{ij}=\pi/2$ and $n_F=30$ (lowest panel). 
Solid lines are the approximated expression for $\vert I_{ij}\vert$ given in Eq.~\eqref{i1}.}\label{fig10}
\end{center}
\end{figure}

We calculate the tunneling matrix element given in Eq.~\eqref{i0} for parallel and highly skewed (almost perpendicular)
configurations of the cylinders. Setting $\gamma_{ij}=0$ in Eq.~\eqref{i0}, and using the wave function for the motion
along the cylinder axis given in Eq.~\eqref{f}, we find:
\begin{align}
\label{iapp1}
I_{ij}^\parallel=&\frac{LK_0(\Delta_{ij}/\xi)}{K_0(R/\xi)^2D}\int_{-\infty}^{+\infty}\! dz\, f_{k_F}(z-z_i)f_{k_F}(z\cos\gamma_{ij}-z_j)\nonumber \\
=&\frac{K_0(\Delta_{ij}/\xi)}{K_0(R/\xi)^2D}\left[(L-\vert z_{ij}\vert)\cos(k_Fz_{ij})+\frac{\sin(k_F\vert z_{ij}\vert)}{k_F}\right],
\end{align} 
where $z_{ij}=z_i-z_j$ and $k_F=\pi n_F/L$, with $n_F$ integer and positive. 
The quantity $\vert I_{ij}^\parallel\vert=\sqrt{I_{ij}^\parallel I_{ji}^\parallel}$ displays an oscillating behavior as a function of the 
misalignment $z_{ij}$, as shown in the top panel of Fig.~\ref{fig10} for $L/D=100$, $\xi/D=0.1$, and for $n_F=15$ and $30$. 
In the same panel we also plot Eq.~\eqref{i1} (solid line) obtained by replacing the wave functions for the motion along the 
cylinder axes with normalized $\theta$-functions. From the figure, we see that Eq.~\eqref{i1}, and so also the analytical formula
given in Eq.~\eqref{iapp}, is approximately equivalent to consider an envelope of the oscillating behavior of the tunneling
matrix element. This equivalence persists also for nonzero angles between the cylinder axes, as shown in the lower panels
of Fig.~\ref{fig10} where Eq.~\eqref{i0} is compared with Eq.~\eqref{i1} for $\gamma_{ij}$ varying from $\pi/100$ to $\pi/2$.
As $\gamma_{ij}$ approaches $\pi/2$, the maxima of Eq.~\eqref{i0} for even and odd values of $n_F$ are respectively
smaller and larger than Eq.~\eqref{i1}, which thus approximately averages the mixture of the tunneling matrix element
for different wave numbers. At exactly $\gamma_{ij}=\pi/2$, $I_{ij}$ for $n_F$ even is identically zero, as seen in the lowest panel of Fig.~\ref{fig10}. To see in more details how $\vert I_{ij}\vert$ behaves for perpendicular or highly skewed configurations
of cylinders we consider Eq.~\eqref{i0} for $\gamma_{ij}\simeq \pi/2$. For $L/D\gg 1$ and $\xi/D\ll 1$ the exponential
decay of $K_0$ limits the  integration over $z$ to $\vert z\vert\lesssim D$ so that $f_{k_F}(z-z_i)f_{k_F}(z\cos\gamma_{ij}-z_j)$
can be approximated by $f_{k_F}(-z_i)f_{k_F}(-z_j)$, which gives \cite{GradshteynBook}:
\begin{align}
\label{iapp2}
I_{ij}^\perp&\simeq\frac{2 F_{ij}}{K_0(R/\xi)^2D}\int_{-\infty}^{+\infty}\!dz
K_0\!\left(\frac{1}{\xi}\sqrt{\Delta_{ij}^2+z^2\sin^2\gamma_{ij}}\right)\nonumber \\
&=\frac{2 \pi\xi F_{ij}}{K_0(R/\xi)^2D}\frac{e^{-\vert\Delta_{ij}\vert/\xi}}{\vert\sin\!\gamma_{ij}\vert}
\simeq 2F_{ij}\frac{e^{-(\vert\Delta_{ij}\vert-D)/\xi}}{\vert\sin\!\gamma_{ij}\vert}
\end{align}
where in the second line we have used the expansion of $K_0$ for large arguments and:
\begin{equation}
\label{iapp3}
F_{ij}=\left\{
\begin{array}{cc}
\cos(k_F z_i)\cos(k_F z_j), & \textrm{$n_F$ odd}\\
\sin(k_F z_i)\sin(k_F z_j), &  \textrm{$n_F$ even}
\end{array}\right.
\end{equation}
Setting $\gamma_{ij}=\pi/2$ and $z_i=0$, we see from Eqs.~\eqref{iapp2} and \eqref{iapp3} that $I_{ij}^\perp=0$ 
for $n_F$ even, while $\vert I_{ij}^\perp\vert \simeq 2\vert\cos(k_F z_j)\vert e^{-(\vert\Delta_{ij}\vert-D)/\xi}$ for $n_F$ odd,
which explains the result shown in the lowest panel of Fig.~\ref{fig10}.


\begin{thebibliography}{50}
\bibitem{notealign} 
For perfectly aligned rods we use $F(\gamma)=\frac{1}{2}\delta(\gamma)+\frac{1}{2}\delta(\gamma-\pi)$
to find $R(1,a)=1/a=L^2/(2\pi\xi D)$.
\end{thebibliography}
\end{document}